\documentclass[showpacs,prd,twocolumn,floatfix,nofootinbib,preprintnumbers,superscriptaddress]{revtex4-1}
\usepackage{booktabs}
\usepackage{amsmath}
\usepackage{siunitx}
\usepackage{url}
\usepackage{longtable}
\usepackage{multirow}
\usepackage{graphicx}
\usepackage{verbatim}
\usepackage{ulem}
\newcommand{\onu}{\overline{\nu}}

\begin{document}

\title{Form factor ratios for $B_s \rightarrow K  \; \ell \, \nu$ and $B_s \rightarrow D_s \; \ell \, \nu$ semileptonic decays and $|V_{ub}/V_{cb}|$}

\author{Christopher J.~Monahan}
\email[e-mail: ]{cjm373@uw.edu}
\affiliation{Institute for Nuclear Theory, University of Washington,
Seattle, Washington 98195-1550, USA}
\author{Chris M.~Bouchard}
\affiliation{School of Physics and Astronomy, University of Glasgow, Glasgow 
G12 
8QQ, United Kingdom}
\author{{G.~Peter} Lepage}
\affiliation{Laboratory of Elementary Particle Physics,
Cornell University, Ithaca, New York 14853, USA}
\author{Heechang Na} 
\affiliation{Ohio Supercomputer Center, 1224 Kinnear Road, Columbus, Ohio 
43212, USA}
\author{Junko Shigemitsu}
\affiliation{Department of Physics,
The Ohio State University, Columbus, Ohio 43210, USA}

\collaboration{HPQCD Collaboration}
\noaffiliation
\date{\today} 


\begin{abstract}
We present a lattice quantum chromodynamics determination of the ratio of the scalar and vector form factors for two semileptonic decays of the $B_s$ meson: $B_s \rightarrow K \ell \nu$ and $B_s \rightarrow D_s \ell \nu$. In conjunction with future experimental data, our results for these correlated form factors will provide a new method to extract $|V_{ub}/V_{cb}|$, which may elucidate the current tension between exclusive and inclusive determinations of these Cabibbo-Kobayashi-Maskawa mixing matrix parameters. In addition to the form factor results, we determine the ratio of the differential decay rates, and forward-backward and polarization asymmetries, for the two decays.
\end{abstract}

\preprint{INT-PUB-18-046}

\maketitle


\section{Introduction}

Semileptonic decays of heavy mesons provide stringent tests of the standard model of particle physics and
opportunities to observe signals of new physics. In particular, experimental measurements of $B$ decays have
highlighted a number of deviations from standard model expectations. These discrepancies include
$R(D^{(\ast)})$, the ratio of the branching fraction of the $B\to D^{(\ast)}\tau\nu$ and $B\to D^{(\ast)}e/\mu\nu$ decays, $R_{K^{(\ast)}}$, the ratio of the branching fraction of the $B \to K^{(\ast)}\mu^+\mu^-$ and $B \to K^{(\ast)}e^+e^-$ decays, and the long-standing tension between inclusive and exclusive determinations of the Cabibbo-Kobayashi-Maskawa (CKM) mixing matrix elements $|V_{ub}|$ and $|V_{cb}|$. Although none of these differences are conclusive evidence of new physics effects, the cumulative weight of
these tensions suggest a hint of new physics.

The ratio $|V_{ub}/V_{cb}|$, which enters into the length of the side of the CKM unitarity triangle opposite the precisely-determined angle $\beta$, is a central input into tests of CKM unitarity. Both $|V_{ub}|$ and $|V_{cb}|$ have been determined through measurements of multiple exclusive mesonic decay channels \cite{pdg18,Amhis:2016xyh}, primarily $B \to \pi \ell \onu_\ell$ \cite{Aubert:2006ry,Hokuue:2006nr,Adam:2007pv,Gray:2007pw,Aubert:2008bf,delAmoSanchez:2010af,delAmoSanchez:2010zd,Ha:2010rf,Sibidanov:2013rkk} and $B \to D^{(\ast)} \ell \onu_\ell$ respectively \cite{Buskulic:1996yq,Abbiendi:2000hk,Abreu:2001ic,Abdallah:2004rz,Adam:2002uw,Aubert:2007rs,Aubert:2007qs,Aubert:2008yv,Dungel:2010uk}, although other channels are also used \cite{Aubert:2003zd,Schwanda:2004fa,Aubert:2008ct,Lees:2012mq,Lees:2013gja}. The $B_s\to K$ decay has generally received less theoretical attention than the corresponding $B$ decay, largely due to the absence of experimental data, although this channel has been studied on the lattice in \cite{Bahr:2014iqa,Flynn:2015mha}, and using other theoretical approaches \cite{Meissner:2013pba}, including light cone sum rules \cite{Duplancic:2008tk,Khodjamirian:2017fxg}, perturbative QCD \cite{Wang:2012ab,Xiao:2014ana} and QCD-inspired models \cite{Cheng:2003sm,Lu:2007sg,Verma:2011yw,Faustov:2013ima,Kang:2018jzg}. Form factors for both $B \to \pi \ell \onu_\ell$  and $B \to D^{(\ast)} \ell \onu_\ell$ decays have been calculated by several lattice groups \cite{Dalgic:2006dt,Colquhoun:2015mfa,Bailey:2015tia,Flynn:2015mha,Bailey:2014tva,Na:2015kha,Lattice:2015rga,Harrison:2017fmw} and using light cone sum rules \cite{Ball:2004ye,Duplancic:2008ix,Khodjamirian:2011ub,Bharucha:2012wy,Imsong:2014oqa,Bigi:1994ga,Kapustin:1996dy,Gambino:2010bp,Gambino:2012rd}, which provide complementary coverage of different kinematic regions. The leptonic decay $B\to \tau \onu$ provides an alternative method to extract $|V_{ub}|$, but this approach is limited by current experimental uncertainties \cite{pdg18}.  Most recently, the ratio $|V_{ub}/V_{cb}|$ was determined by the LHCb collaboration through the ratio of the baryonic decays $\Lambda_b^0 \to \Lambda_c^+\mu\onu$ and $\Lambda_b^0 \to p\mu\onu$ \cite{Aaij:2015bfa,Aaij:2017svr}, using form factors determined with lattice QCD \cite{Detmold:2015aaa}. Inclusive determinations of $|V_{ub}|$ differ from the value extracted from exclusive decays at the level of approximately three standard deviations.

Here we undertake a correlated study of the form factors for the $B_s\to K\ell \onu_\ell$ and $B_s\to D_s\ell \onu_\ell$ decays, which, in conjunction with anticipated experimental results from the LHCb Collaboration, will provide a new method to determine the ratio $|V_{ub}/V_{cb}|$. We perform a chiral-continuum-kinematic fit to the scalar and vector form factors for both the $B_s\to K\ell \onu_\ell$ \cite{Bouchard:2014ypa} and $B_s\to D_s\ell \onu_\ell$ decays \cite{Monahan:2017uby}, to determine the correlated form factors over the full range of momentum transfer. Using the ratio of the form factors significantly reduces the largest systematic uncertainty at large values of the momentum transfer, which stems from the perturbative matching of lattice nonrelativistic QCD (NRQCD) currents to continuum QCD. We use our form factor results to predict several phenomenological ratios, including the differential branching fractions, and the forward-backward and polarization asymmetries.

We briefly summarize the details of the lattice calculations used in the analyses of \cite{Bouchard:2014ypa,Monahan:2017uby} in Sec.~\ref{sec:lattsetup} and the corresponding form factor results in Sec.~\ref{sec:corrfits}. We then present our new chiral-continuum-kinematic extrapolation in Sec.~\ref{sec:zexp}, and our phenomenological predictions in Sec.~\ref{sec:results}, before summarizing in Sec.~\ref{sec:summary}. We provide further details of the input two-point correlator data in Appendix \ref{app:twopt} and details required to reconstruct our chiral-continuum-kinematic fit in Appendix \ref{sec:ffdetails}.

\section{\label{sec:lattsetup}Ensembles, currents and correlators}

Our determination of the ratio of the form factors for the exclusive $B_s \rightarrow X_s \ell \nu$ 
semileptonic decays closely parallels the analyses presented in 
\cite{Bouchard:2014ypa,Na:2015kha,Monahan:2017uby}. Throughout this work, we use $X_s$ to represent a $K$ or $D_s$ meson. We use the two- and three-point correlator data presented in \cite{Bouchard:2014ypa,Monahan:2017uby} to perform a simultaneous, correlated fit of the form factors for both $B_s \rightarrow K \ell \nu$ and $B_s \rightarrow D_s \ell \nu$ decays. In this section we outline the details of the ensembles, reproduce the form factor results for convenience, and refer the reader to \cite{Bouchard:2014ypa,Na:2015kha,Monahan:2017uby} for details of the correlator analysis.

We use five gauge ensembles with $n_f = 2+1$ flavors of AsqTad sea quarks generated by 
the MILC Collaboration \cite{Bazavov:2009bb}, including three 
``coarse'' (with 
lattice spacing $a \approx \SI{0.12}{fm}$) and two ``fine'' (with $a \approx 
\SI{0.09}{fm}$) ensembles.
\begin{table}
\caption{\label{tab:milc}
Details of three ``coarse'' and two ``fine''  $n_f = 2 + 1$ MILC 
ensembles used in the determination of the scalar and vector form
factors.
}
\begin{ruledtabular}
\begin{tabular}{cccccc}
Set &  $r_1/a$ & $m_l/m_s$ (sea)   &  $N_{\mathrm{conf}}\,(K/D_s)$&
$N_{\mathrm{tsrc}}$ & $L^3 \times N_t$ \\
\vspace*{-10pt}\\
\hline 
\vspace*{-6pt}\\
C1  & 2.647 & 0.005/0.050   & 1200/2096  &  2/4 & $24^3 \times 64$ \\
C2  & 2.618 & 0.010/0.050  & 1200/2256   & 2/2 & $20^3 \times 64$ \\
C3  & 2.644 & 0.020/0.050  & 600/1200  & 2/2 & $20^3 \times 64$ \\
F1  & 3.699 & 0.0062/0.031  & 1200/1896  & 4/4  & $28^3 \times 96$ \\
F2  & 3.712 & 0.0124/0.031  & 600/1200  & 4/4 & $28^3 \times 96$ \\
\end{tabular}
\end{ruledtabular}
\end{table}
We summarize these ensembles in Table \ref{tab:milc} and tabulate the corresponding light pseudoscalar masses, for 
both AsqTad and HISQ valence quarks, in Table \ref{tab:deltapi}.
\begin{table*}
\caption{\label{tab:deltapi}
Light meson masses on MILC ensembles for both AsqTad \cite{Bazavov:2009bb} and HISQ 
valence quarks \cite{Bouchard:2014ypa}. In the final column we list the finite volume corrections to chiral logarithms from staggered perturbation theory \cite{Bernard:2001yj}, for each ensemble.}
\begin{ruledtabular}
\begin{tabular}{cccccccc}
Set & $aM_\pi^{\mathrm{AsqTad}}$   & $aM_\pi^{\mathrm{HISQ}}$ & 
$aM_K^{\mathrm{AsqTad}}$ & $aM_K^{\mathrm{HISQ}}$ & $aM_{\eta_s}^{\mathrm{HISQ}}$ & $aM_{D_s}^{\mathrm{HISQ}}$ & $\delta_{\mathrm{FV}}$\\
\vspace*{-10pt}\\
\hline 
\vspace*{-6pt}\\
C1 & 0.15971(20) & 0.15990(20) & 0.36530(29) & 0.31217(20) & 0.41111(12) & 1.18755(22) & 0.053647 \\
C2 & 0.22447(17) & 0.21110(20) & 0.38331(24) & 0.32851(48) & 0.41445(17) & 1.20090(30) & 0.030760 \\
C3 & 0.31125(16) & 0.29310(20) & 0.40984(21) & 0.35720(22) & 0.41180(23) & 1.19010(33) & 0.003375 \\
F1 & 0.14789(18) & 0.13460(10) & 0.25318(19) & 0.22855(17) & 0.294109(93) & 0.84674(12) & 0.059389 \\
F2 & 0.20635(18) & 0.18730(10) & 0.27217(21) & 0.24596(14) & 0.29315(12) & 0.84415(14) & 0.007567 \\
\end{tabular}
\end{ruledtabular}
\end{table*}

In Table \ref{tab:valq} we list the valence quark masses for the NRQCD 
bottom quarks and HISQ charm quarks \cite{Na:2012kp,Bouchard:2014ypa}. For 
completeness and ease of reference, we include 
both the tree-level wave function renormalization for the massive HISQ quarks 
\cite{Monahan:2012dq} and the spin-averaged $\Upsilon$ mass, corrected for 
electroweak effects, determined in \cite{Na:2012kp}.
\begin{table}
\caption{\label{tab:valq}
Valence quark masses $a m_b$ for NRQCD bottom quarks and 
 $a m_s$ and $a m_c$ for HISQ  strange
 and charm quarks.  The fifth
column gives 
$Z_2^{(0)}(a m_c)$, the tree-level wave function renormalization 
constant for massive (charm) HISQ quarks. The sixth column lists the values of 
the spin-averaged $\Upsilon$ mass, corrected for electroweak effects.}
\begin{ruledtabular}
\begin{tabular}{cccccc}
Set & $a m_b$   & $a m_s$ & $a m_c$ & $  Z_2^{(0)}(a m_c) $ & 
$aE_{b\overline{b}}^{\mathrm{sim}}$  \\
\vspace*{-10pt}\\
\hline 
\vspace*{-6pt}\\
C1 & 2.650 & 0.0489 & 0.6207 & 1.00495618 & 0.28356(15) \\
C2 & 2.688 & 0.0492 & 0.6300 & 1.00524023 & 0.28323(18) \\
C3 & 2.650 & 0.0491 & 0.6235 & 1.00504054 & 0.27897(20) \\
F1 & 1.832 & 0.0337 & 0.4130 & 1.00103879 & 0.25653(14) \\
F2 & 1.826 & 0.0336 & 0.4120 & 1.00102902 & 0.25558(28) \\
\end{tabular}
\end{ruledtabular}
\end{table}

The scalar, $f_0^{(X_s)}(q^2)$, and vector, $f_+^{(X_s)}(q^2)$, form factors that characterize 
the $B_s \rightarrow X_s$ semileptonic decays are defined by the matrix element
\begin{align}
\langle  X_s{} & (p_{X_s})| V^\mu | B_s(p_{B_s}) \rangle = f_0^{(X_s)}(q^2) 
\frac{M_{B_s}^2-M_{X_s}^2}{q^2}q^\mu \nonumber\\
{} & + f_+^{X_s}(q^2) \left[
p_{B_s}^\mu + p_{X_s}^\mu - \frac{M_{B_s}^2-M_{X_s}^2}{q^2}q^\mu
\right],
\end{align}
where $V^\mu$ is a flavor-changing vector current and the momentum transfer is $q^\mu = p_{B_s}^\mu - p_{X_s}^\mu$. On the lattice
it is more convenient to work with the form factors $f_\parallel^{(X_s)}$ and 
$f_\perp^{(X_s)}$, which are given in terms of the scalar and vector form factors by
\begin{align}
f_+^{(X_s)}(q^2) = {} & \frac{1}{\sqrt{2M_{B_s}}}\Big[f_\parallel^{(X_s)}(q^2) 
\nonumber\\
{} & \qquad
+ 
(M_{B_s}-E_{X_s})f_\perp^{(X_s)}(q^2)\Big], \\
f_0^{(X_s)}(q^2) = {} & 
\frac{\sqrt{2M_{B_s}}}{M_{B_s}^2-M_{X_s}^2}\bigg[(M_{B_s}-E_{X_s} 
)f_\parallel^{(X_s)}(q^2) \nonumber\\
{} & \qquad + 
(E_{X_s}^2-M_{X_s}^2)f_\perp^{(X_s)}(q^2)\bigg].
\end{align}
Here $E_{X_s}$ is the energy of the $X_s$ meson in the rest frame of 
the $B_s$ meson. We work in the rest frame of the $B_s$ meson 
and throughout the rest of this work the spatial momentum, $\vec{p}$, denotes the momentum of 
the $X_s$ meson.

NRQCD is an effective theory for heavy quarks and results determined using 
lattice
NRQCD must be matched to full QCD to make contact with experimental data. We 
match the bottom-charm currents, $J_\mu$, at one loop in perturbation theory 
through ${\cal O}(\alpha_s, \Lambda_{\mathrm{QCD}}/m_b, \alpha_s/(am_b))$, where 
$am_b$ is the bare lattice mass \cite{Monahan:2012dq}. We rescale all currents 
by the nontrivial massive wave function renormalization for the HISQ charm 
quarks, tabulated in Table \ref{tab:valq}, and taken from \cite{Na:2015kha,Monahan:2012dq}.

\begin{figure}
\centering
\caption{\label{fig:3pt}Lattice setup for the three-point correlators. See 
accompanying text for details.}
\includegraphics[width=0.45\textwidth,keepaspectratio]{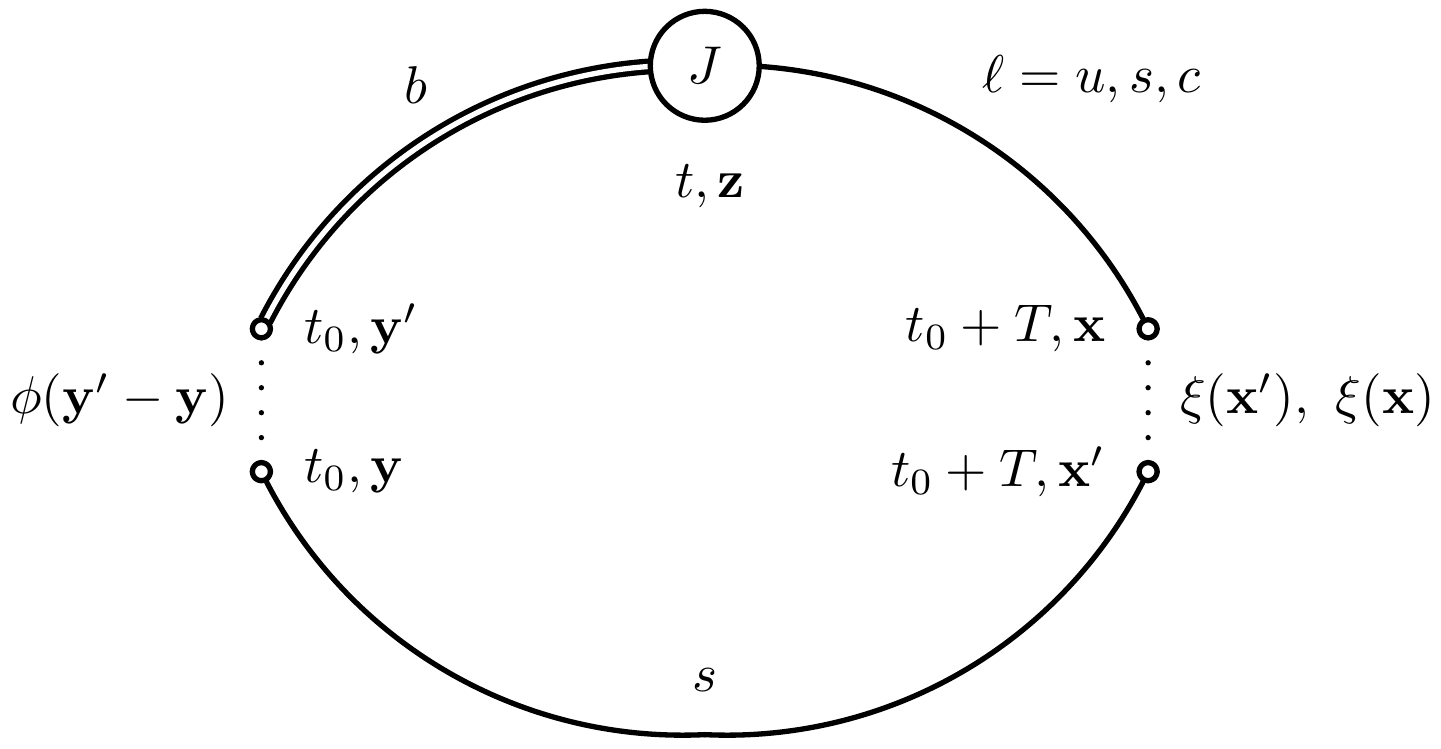}
\end{figure}

The $B_s$ and $X_s$ meson two-point correlators and 
three-point correlators of the NRQCD-HISQ currents, $J_\mu$, were calculated in \cite{Bouchard:2014ypa,Monahan:2017uby}. 
In those calculations, we used smeared heavy-strange 
bilinears to represent the $B_s$ meson and incorporated both delta-function and 
Gaussian smearing, with a smearing radius of $r_0/a = 5$ and $r_0/a = 7$ on the 
coarse and fine ensembles, respectively. The three-point correlators were determined 
with the setup illustrated in Fig.~\ref{fig:3pt}.  The $B_s$ meson is created 
at time $t_0$ and a current $J_\mu$ inserted at time $t$, between $t_0$ 
and $t_0+T$. The $X_s$ meson is then annihilated at time 
$t_0+T$. We used four values of $T$: 12, 13, 14, and 15 on the coarse lattices; 
and 21, 22, 23, and 24 on the fine lattices. We implemented spatial sums at the 
source through the $U(1)$ random wall sources $\xi(x)$ and $\xi(x')$ 
\cite{Na:2010uf} and generated data for four different values of the $X_s$ meson 
momenta, $\vec{p} = 2\pi/(aL)(0,0,0)$, $\vec{p} = 
2\pi/(aL)(1,0,0)$, $\vec{p} = 2\pi/(aL)(1,1,0)$, and $\vec{p} = 
2\pi/(aL)(1,1,1)$, where $L$ is the spatial lattice extent.

\section{\label{sec:corrfits}Correlator and form factor results}

The results for the two- and three-point correlators were determined with a Bayesian multiexponential fitting procedure, based on the 
\verb+PYTHON+ packages \verb+LSQFIT+ \cite{lsqfit} and \verb+CORRFITTER+ 
\cite{corrfitter}. The results are summarized for convenience in Appendix \ref{app:twopt}.

We summarize the final results for the form factors, $f_0(\vec{p})$ and 
$f_+(\vec{p})$, for each ensemble and $X_s$ momentum in Tables 
\ref{tab:ff_k} and 
\ref{tab:ff_ds}.
\begin{table*}
\caption{\label{tab:ff_k}
Final results for the form factors $f_0^{(K)}(\vec{p})$ and $f_+^{(K)}(\vec{p})$. Data reproduced from Table II of \cite{Bouchard:2014ypa}.}
\begin{ruledtabular}
\begin{tabular}{cccccccc}
Set & $f_0^{(K)}(0,0,0)$   & $f_0^{(K)}(1,0,0)$ & $f_0^{(K)}(1,1,0)$ & $  
f_0^{(K)}(1,1,1)$ & $f_+^{(K)}(1,0,0)$   & $f_+^{(K)}(1,1,0)$ & $f_+^{(K)}(1,1,1)$ \\
\vspace*{-10pt}\\
\hline 
\vspace*{-8pt}\\
C1 & 0.8244(23) & 0.7081(27) & 0.6383(30) & 0.5938(41) & 2.087(16) & 1.657(14) & 1.378(13) \\
C2 & 0.8427(25) & 0.6927(35) & 0.6036(49) & 0.536(12) & 1.880(12) & 1.412(16) & 1.142(33) \\
C3 & 0.8313(29) & 0.6953(33) & 0.6309(30) & 0.5844(46) & 1.773(11) & 1.4212(84) & 1.184(10) \\
F1 & 0.8322(25) & 0.6844(35) & 0.5994(43) & 0.5551(56) & 1.878(13) & 1.385(12) & 1.158(13) \\
F2 & 0.8316(27) & 0.6915(38) & 0.6199(43) & 0.5563(61) & 1.834(14) & 1.396(10) & 1.163(14) \\
\end{tabular}
\end{ruledtabular}
\end{table*}
\begin{table*}
\caption{\label{tab:ff_ds}
Final results for the form factors $f_0^{(D_s)}(\vec{p})$ and $f_+^{(D_s)}(\vec{p})$. Data reproduced from Tables VI and VII of \cite{Monahan:2017uby}.}
\begin{ruledtabular}
\begin{tabular}{cccccccc}
Set & $f_0^{(D_s)}(0,0,0)$   & $f_0^{(D_s)}(1,0,0)$ & $f_0^{(D_s)}(1,1,0)$ & $  
f_0^{(D_s)}(1,1,1)$ & $f_+^{(D_s)}(1,0,0)$   & $f_+^{(D_s)}(1,1,0)$ & $f_+^{(D_s)}(1,1,1)$ \\
\vspace*{-10pt}\\
\hline 
\vspace*{-8pt}\\
C1 & 0.8885(11) & 0.8754(14) & 0.8645(13) & 0.8568(13) & 1.1384(35) & 1.1081(20) & 1.0827(21) \\
C2 & 0.8822(13) & 0.8663(15) & 0.8524(16) & 0.8418(18) & 1.1137(29) & 1.0795(22) & 1.0470(21) \\
C3 & 0.8883(13) & 0.8723(16) & 0.8603(16) & 0.8484(21) & 1.1260(34) & 1.0912(24) & 1.0552(28) \\
F1 & 0.90632(98) & 0.8848(13) & 0.8674(13) & 0.8506(17) & 1.1453(29) & 1.0955(24) & 1.0549(24) \\
F2 & 0.9047(12) & 0.8855(16) & 0.8667(15) & 0.8487(19) & 1.1347(42) & 1.0905(26) & 1.0457(33) \\
\end{tabular}
\end{ruledtabular}
\end{table*}
For more details, see \cite{Bouchard:2014ypa,Na:2015kha,Monahan:2017uby}.

\section{\label{sec:zexp}Chiral, continuum and kinematic extrapolations}

Form factors determined from experimental data are functions of a single
kinematic variable, which is typically the momentum transfer, 
$q^2$, or the energy of the mesonic decay product, $E_{X_s}$. Alternatively, the form factors can 
be expressed in terms of the 
$z$-variable,
\begin{equation}\label{eq:zdef}
z(q^2) = \frac{\sqrt{t_+-q^2} - \sqrt{t_+-t_0}}{\sqrt{t_+-q^2}+\sqrt{t_+-t_0}}.
\end{equation}
Here $t_+ = (M_{B_s}+M_{X_s})^2$ and $t_0$ is a free parameter, which we take 
to be $t_0 = (M_{B_s}+M_{X_s})(\sqrt{M_{B_s}}+\sqrt{M_{X_s}})^2$, as in \cite{Bouchard:2014ypa}. This choice minimizes the magnitude of $z$ over the physical range of momentum transfer.
Note that in \cite{Monahan:2017uby} the choice $t_0 = q_{\mathrm{max}}^2= (M_{B_s}-M_{X_s})^2$ was used to ensure consistency with the analysis of 
\cite{Na:2015kha}. We have confirmed that our extrapolation results are independent of our choice of $t_0$, within
fit uncertainties.

Lattice calculations of form factors are necessarily determined at finite lattice spacing,
generally with light quark masses that are heavier than their physical values, and are
thus functions of the lattice spacing and the light quark mass in addition to the momentum
transfer. We remove the lattice spacing and light quark mass dependence of the lattice
results by performing a combined continuum-chiral-kinematic extrapolation, through
the modified $z$-expansion, which was introduced in 
\cite{Na:2010uf,Na:2011mc} and applied to $B_s$ semileptonic decays in 
\cite{Bouchard:2013mia,Bouchard:2013pna,Bouchard:2014ypa,Monahan:2017uby}.

Our chiral-continuum-kinematic extrapolation for the $B_s \to X_s \ell \nu$ decays 
closely parallels those studied in \cite{Bouchard:2014ypa,Na:2015kha,Monahan:2017uby}, 
so here we outline the main components and refer the reader to those references
for details.

The dependence of the form factors on the $z$-variable is expressed through a 
modification of the Bourrely-Caprini-Lellouch (BCL)
parametrization \cite{Bourrely:2008za}
\begin{align}
P_0^{(X_s)}f_0^{(X_s)}(q^2(z)) = {} & \left[1+L^{(X_s)}\right] \nonumber \\
{} & \; \times \sum_{j=0}^{J-1} 
a_j^{(0,X_s)}(m_l,m_l^{\mathrm{sea}},a) 
z^j, \\
P_+^{(X_s)}f_+^{(X_s)}(q^2(z)) = {} & \left[1+L^{(X_s)}\right]\nonumber\\
 \times\sum_{j=0}^{J-1} 
a_j^{(+,X_s)}(m_l,{} &m_l^{\mathrm{sea}},a)
\left[z^j - (-1)^{j-J}\frac{j}{J}z^J\right].
\end{align}
Here the $P_{0,+}$ are Blaschke factors that take into account the effects of 
expected poles above the physical region, 
\begin{equation}
P_{0,+}^{(X_s)}(q^2) = \left(1-\frac{q^2}{\Big(M_{0,+}^{(X_s)}\Big)^2}\right),
\end{equation}
where we take \cite{Gregory:2009hq,Bouchard:2014ypa,Monahan:2017uby}
\begin{align}
M_+^{(K)} = {} & \SI{5.32520(48)}{GeV}, \\
M_0^{(K)} = {} & \SI{5.6794(10)}{GeV}, \\
M_+^{(D_s)} = {} & M_{B_c^\ast} = \SI{6.330(9)}{GeV}, \\
M_0^{(D_s)} = {} & \SI{6.42(10)}{GeV}.
\end{align}
In line with \cite{Bouchard:2014ypa}, we convert these values to lattice units
in the chiral-continuum-kinematic extrapolation, so that the difference between 
the ground state meson masses and these pole masses is fixed in physical units.

The functions $L^{(X_s)}$ incorporate the chiral logarithmic corrections, which
are fixed by hard pion chiral perturbation theory \cite{Bijnens:2010ws,Bijnens:2010jg} for the $B_s\to K$ decay
\begin{align}
L^K = {} & -\frac{3}{8}x_\pi(\log x_\pi + \delta_{\mathrm{FV}}) - 
\frac{1+6g^2}{4}x_K \log x_K \nonumber \\
{} & \qquad - \frac{1+12g^2}{24}x_\eta \log x_\eta.
\end{align}
Here $g^2 = 0.51(20)$, $\delta_{\mathrm{FV}}$ are finite volume corrections given in Table \ref{tab:deltapi}, we define
\begin{align}
x_{\pi,K,\eta,\eta_s} = {} & \frac{M_{\pi,K,\eta,\eta_s}^2}{(4\pi f_\pi)^2}, \\
\delta x_{\pi,K} = {} & \frac{(M_{\pi,K}^{\mathrm{AsqTad}})^2 - 
(M_{\pi,K}^{\mathrm{HISQ}})^2
}{(4\pi f_\pi)^2},  \\
\delta x_{\eta_s} = {} & \frac{(M_{\eta_s}^{\mathrm{HISQ}})^2 - 
(M_{\eta_s}^{\mathrm{phys.}})^2
}{(4\pi f_\pi)^2}, 
\end{align}
and $M_\eta^2 = (M_\pi^2+2M_{\eta_s}^2)/3$. We tabulate the meson masses required to calculate $\delta x_{\pi,K,\eta_s}$ in Table \ref{tab:deltapi}. For the $B_s\to D_s$ decay, the chiral logarithmic corrections cannot be factored
out in the $z$-expansion \cite{Bijnens:2010jg} and therefore we follow \cite{Na:2015kha,Monahan:2017uby}
and fit the logarithmic dependence by introducing corresponding fit parameters
in the expansion coefficients 
$a_j^{(0,+,D_s)}$. In other words, we take
\begin{equation}
L^{(D_s)} = 0,
\end{equation}
and introduce an appropriate fit parameter, $c_j^{(2)}$, in the corresponding fit function, Eq.~\eqref{eq:DjDs}.

The expansion coefficients 
$a_j^{(0,+,X_s)}$ include 
lattice spacing and quark mass dependence and can be written as
\begin{equation}
a_j^{(0,+,X_s)}(m_l,m_l^{\mathrm{sea}},a)  = 
\widetilde{a}_j^{(0,+,X_s)}\widetilde{D}_j^{(0,+,X_s)}(m_l,m_l^{\mathrm{sea}},a),
\end{equation}
where the $\widetilde{D}_j^{(0,+,X_s)}$ include all lattice artifacts. Suppressing the $0,+$ superscripts for clarity, these coefficients are given by \cite{Bouchard:2014ypa}
\begin{align}\label{eq:DjK}
\widetilde{D}_j^{(K)} = {} & 1 + c_j^{(1)} x_\pi+ 
d_j^{(1)}\left(\frac{\delta x_\pi}{2} + \delta x_K\right) + 
d_j^{(2)}\delta x_{\eta_s}\nonumber\\
{} & \qquad    + e_j^{(1)} \left(\frac{aE_K}{\pi}\right)^2 + e_j^{(2)}
\left(\frac{aE_K}{\pi}\right)^4 \nonumber\\
{} & \qquad + f_j^{(1)} \left(\frac{a}{r_1}\right)^2 + f_j^{(2)} \left(\frac{a}{r_1}\right)^4,
\end{align}
and \cite{Monahan:2017uby}
\begin{align}\label{eq:DjDs}
\widetilde{D}_j^{(D_s)} = {} & 1 + c_j^{(1)} x_\pi+ c_j^{(2)} x_\pi 
\log(x_\pi) \nonumber\\
{} & \qquad + 
d_j^{(1)}\left(\frac{\delta x_\pi}{2} + \delta x_K\right)  + 
d_j^{(2)}\delta x_{\eta_s}
\nonumber\\
{} & \qquad + e_j^{(1)} \left(\frac{aE_{D_s}}{\pi}\right)^2 + e_j^{(2)}
\left(\frac{aE_{D_s}}{\pi}\right)^4 \nonumber\\
{} & \qquad + m_j^{(1)} (am_c)^2 + m_j^{(2)} (am_c)^4.
\end{align}
Here the $c_j^{(i)}$, $d_j^{(i)}$, $e_j^{(i)}$, $f_j^{(i)}$, and $m_j^{(i)}$ are fit 
parameters, 
along with the $\widetilde{a}_j^{(0,+)}$. We incorporate light- and heavy-quark mass dependence in the discretization coefficients $f_j^{(i)}$ by replacing
\begin{align}
f_j^{(i)}\to {} & f_j^{(i)}(1+l_j^{(1,i)}x_\pi + l_j^{(2,i)}x_\pi^2) \nonumber \\
{} & \qquad \times (1+h_j^{(1,i)}\delta m_b + h_j^{(2,i)}(\delta m_b)^2),
\end{align}
where $\delta m_b = am_b - 2.26$ \cite{Bouchard:2014ypa} and is chosen to minimize the magnitude of $\delta m_b$, such that $-0.4<\delta m_b < 0.4$. Here $x_\pi$ captures sea pion mass dependence and is determined from the AsqTad pion mass \cite{Bazavov:2009bb}.

The actions we use are highly improved and ${\cal O}(a^2)$ tree-level lattice artifacts have been removed. The ${\cal O}(\alpha_s a^2 )$ and ${\cal O}(a^4)$ corrections are dominated by powers of $(am_c)$ and $(aE_{X_s})$, rather than those of the spatial momenta $(ap_i)$.  Thus, we do not incorporate terms involving hypercubic invariants constructed from the spatial momentum $ap_i$ \cite{Lubicz:2016wwx}. 

We follow \cite{Bouchard:2014ypa,Monahan:2017uby} and impose the kinematic constraint $f_0(0) = f_+(0)$ analytically for the $B_s\to K$ decay, and as a data point for the $B_s\to D_s$ channel. To incorporate the systematic uncertainty associated with truncation of the 
perturbative current-matching procedure at ${\cal O}(\alpha_s, \Lambda_{\mathrm{QCD}}/m_b, 
\alpha_s/(am_b))$, we introduce fit parameters $m_\parallel$ and $m_\perp$, 
with central value zero and width $\delta m_{\parallel,\perp}$ and re-scale the 
form factors, $f_\parallel$ and $f_\perp$ according to
\begin{equation}
f_{\parallel,\perp} \rightarrow (1+m_{\parallel,\perp})f_{\parallel,\perp}.
\end{equation}
We take $\delta m_{\parallel,\perp} = 0.04$. We refer to this fit Ansatz, including terms 
up to $z^3$ in the modified $z$-expansion, as the 
``standard extrapolation.''

To test the convergence of our fit Ansatz and ensure we have included a sufficient number of terms in the modified $z$-expansion, we modify the fit Ansatz in the following ways:
\begin{enumerate}
\item include terms up to $z^2$ in the $z$-expansion;
\item include terms up to $z^4$ in the $z$-expansion;
\item include discretization terms up to $(am_c)^2$;
\item include discretization terms up to $(am_c)^6$;
\item include discretization terms up to $(a/r_1)^2$;
\item include discretization terms up to $(a/r_1)^6$;
\item include discretization terms up to $(aE_K/\pi)^2$;
\item include discretization terms up to $(aE_K/\pi)^6$;
\item include discretization terms up to $(aE_{D_s}/\pi)^2$;
\item include discretization terms up to $(aE_{D_s}/\pi)^6$;
\end{enumerate}
\begin{figure}
\centering
\caption{\label{fig:deltaz1}Comparison of the convergence tests of the ``standard 
extrapolation'' fit Ansatz. The top panel shows the $\chi^2/\mathrm{dof}$ for each test,
normalized by the $\chi^2/\mathrm{dof}$ for the standard extrapolation. The lower panel
shows the fit results for the form 
factor ratio $f_0^{(K)}/f_0^{(D_s)}$ at $q^2=0$.
The test 
numbers labeling the horizontal axis 
correspond to the modifications listed in the text. The first data point, 
the purple square, is the 
``standard extrapolation'' 
fit result, which is also represented by the purple shaded 
band.}
\includegraphics[width=0.44\textwidth,keepaspectratio]{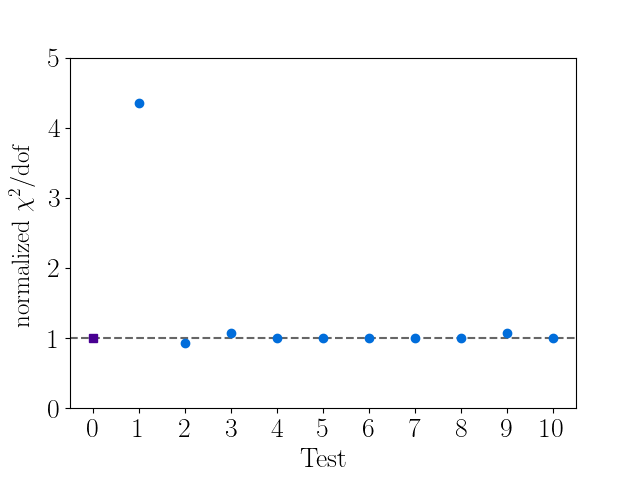}\\
\includegraphics[width=0.44\textwidth,keepaspectratio]{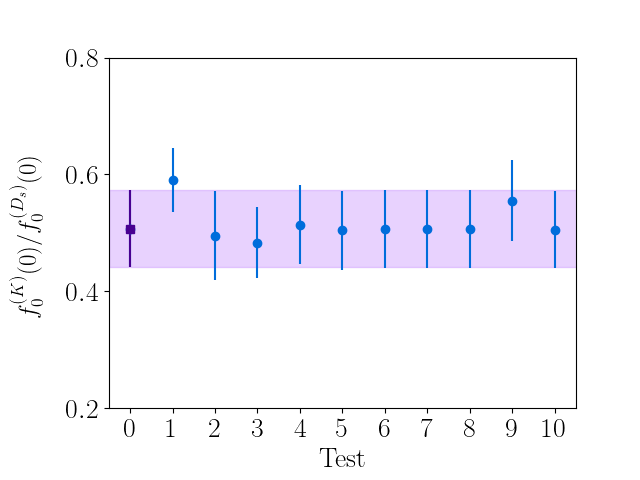}
\end{figure}
We show the results of these modifications in Fig.~\ref{fig:deltaz1}, where we label the standard fit Ansatz as ``Test 0''. These tests demonstrate the stability of the standard fit Ansatz; adding higher order terms does not alter the fit results or improve the goodness-of-fit. 

\begin{figure}
\centering
\caption{\label{fig:deltaz2}Analogous to Fig.~\ref{fig:deltaz1}, but for stability tests labeled by ``i.'' to ``xii.'' in the text. Details provided in the caption of Fig.~\ref{fig:deltaz1}.}
\includegraphics[width=0.44\textwidth,keepaspectratio]{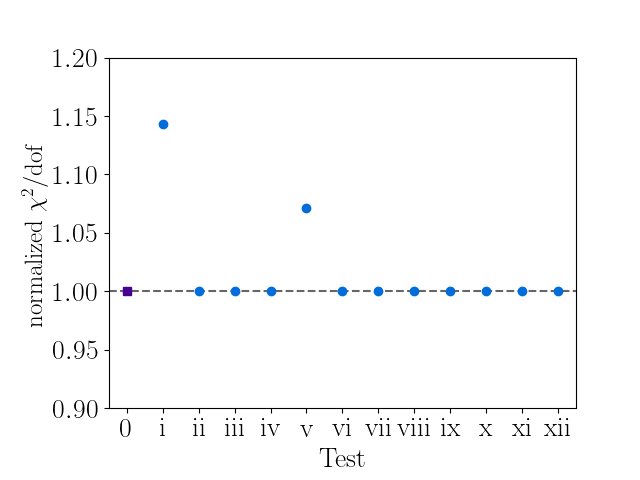}\\
\includegraphics[width=0.44\textwidth,keepaspectratio]{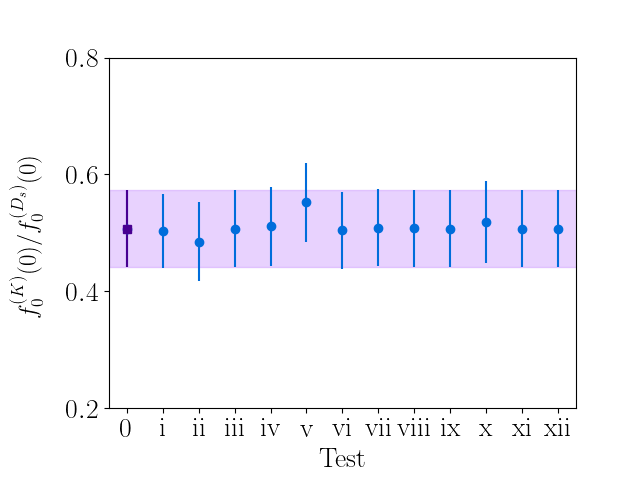}
\end{figure}

We also study the stability of the fit with respect to the following variations:
\renewcommand{\theenumi}{\roman{enumi}}
\begin{enumerate}
\item omit the $x_\pi \log(x_\pi)$ term;
\item omit the light quark mass-dependent discretization terms from the $f_j^{(i)}$ coefficients;
\item add strange quark mass-dependent discretization terms to the $f_j^{(i)}$ coefficients;
\item omit the $am_b$-dependent discretization terms from the $f_j^{(i)}$ coefficients;
\item omit sea- and valence-quark mass difference, $d_j^{(1)}$;
\item omit the strange quark mass mistuning, $d_j^{(2)}$;
\item omit finite volume effects;
\item add light-quark mass dependence to the $m_j^{(i)}$ fit parameters;
\item add strange-quark mass dependence to the $m_j^{(i)}$ fit parameters;
\item add bottom-quark mass dependence to the $m_j^{(i)}$ fit parameters;
\item incorporate a 2\% uncertainty for higher-order matching contributions;
\item incorporate a 5\% uncertainty for higher-order matching contributions;
\end{enumerate}
We show the results of these stability tests in Fig.~\ref{fig:deltaz2}. Test 0 represents the standard fit Ansatz. Taken together, these plots 
demonstrate that the fit has converged with respect to a 
variety of modifications of the chiral-continuum-kinematic extrapolation 
Ansatz.

\section{\label{sec:results}Results}

\subsection{Form factor ratios}

Our final results, from a simultaneous fit to both decay channels, for the ratio of form factors at zero momentum transfer are
\begin{equation}
\frac{f_0^{(K)}(0)}{f_0^{(D_s)}(0)} = 0.507(66),\label{eq:ffratios1}
\end{equation}
where the uncertainties account for correlations between the form factor results
for each decay channel. The corresponding results for the individual form factors at zero momentum transfer
are $ f_0^{(K)}(0) = 0.341(42)$ and $f_0^{(D_s)}(0) = 0.661(42)$, in good agreement with the results of \cite{Bouchard:2014ypa} and \cite{Monahan:2017uby}, respectively. The result in Eq.~\eqref{eq:ffratios1} is in good agreement with, but with significantly reduced uncertainties, the value obtained assuming uncorrelated uncertainties between the results of \cite{Bouchard:2014ypa,Monahan:2017uby}: $f_0^{(K)}(0)/f_0^{(D_s)}(0) = 0.323(63)/0.656(31) = 0.492(99)$. 

We obtain a reduced 
$\chi^2$ of $\chi^2/\mathrm{dof} = 1.3$ with 71 degrees of freedom (dof), with 
a quality factor of $Q=0.011$. The $Q$-value (or $p$-value) corresponds to the 
probability that the $\chi^2/\mathrm{dof}$ from the fit could have been larger, 
by chance, assuming the data are all Gaussian and consistent with each other. The simultaneous fit ensures that the uncertainties
associated with the perturbative matching procedure for the heavy-light currents largely cancel in the form factor ratio. This
can be seen by comparing the error budget contribution from perturbative matching in Table \ref{tab:ffratioerrors}, with the individual fits,
for which the perturbative truncation uncertainty was the second-largest source of uncertainty. The uncertainties in our ratio results
are dominated by the $B_s\to K$ channel, which has fewer statistics and a larger extrapolation uncertainty, because, in the region of momentum transfer reported here, $0--12.5$ GeV$^2$, the corresponding form factors are extrapolated further from the region in which we have lattice results.

We tabulate our choice 
of priors and the fit results in the Appendix, where we provide the 
corresponding $z$-expansion coefficients and their correlations. Following \cite{Monahan:2017uby}, based on the 
earlier work of \cite{Na:2010uf,Na:2011mc,Na:2015kha}, we split the priors into 
three groups. Broadly 
speaking, Group I priors includes the 
typical fit parameters, Group II the input lattice scales and masses, 
and Group III priors the inputs from experiment, such as physical meson masses. We plot our final results for the ratios of the form factors, $f_0^{(K)}/f_0^{(D_s)}(q^2)$ and $f_+^{(K)}/f_+^{(D_s)}(q^2)$, as a 
function of the momentum transfer, $q^2$, in Fig.~\ref{fig:ffratioq2}. Details required to reconstruct the fully correlated form factors are given in Appendix \ref{sec:ffdetails}.
\begin{figure}
\centering
\caption{\label{fig:ffratioq2}Chiral and continuum extrapolated form factor ratios, 
$f_0^{(K)}/f_0^{(D_s)}(q^2)$ (upper panel) and $f_+^{(K)}/f_+^{(D_s)}(q^2)$ (lower panel), as a function of the 
momentum transfer, $q^2$. The dashed lines indicate the central values of the extrapolated form factors and the uncertainty bands include all sources of statistical and systematic uncertainty.}
\includegraphics[width=0.44\textwidth,keepaspectratio]{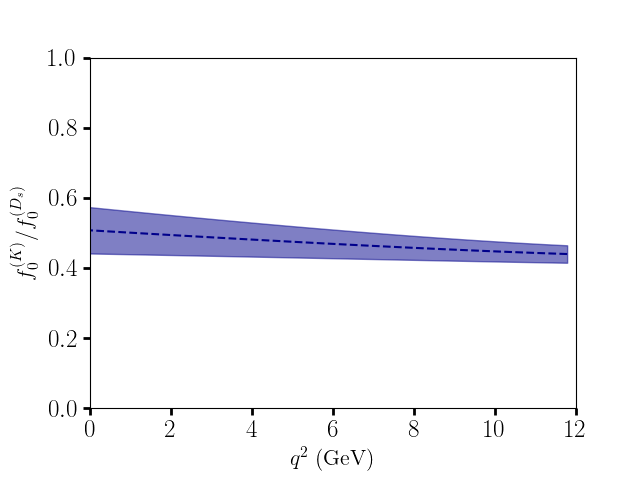}\\
\includegraphics[width=0.44\textwidth,keepaspectratio]{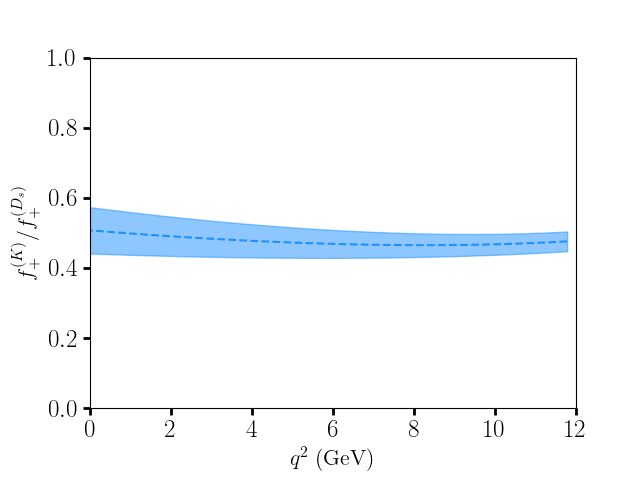}
\end{figure}

\subsection{Form factor error budget}

We tabulate 
the errors in the ratios of the form factors at zero momentum transfer, Eq.~
\eqref{eq:ffratios1}, in Table \ref{tab:ffratioerrors}.
\begin{table}
\caption{\label{tab:ffratioerrors}
Error budget for the form factor ratios at zero momentum transfer, Eq.~\eqref{eq:ffratios1}. 
We describe each source of uncertainty in more detail in the accompanying text.
}
\begin{ruledtabular}
\begin{tabular}{cc}
Type & Partial uncertainty (\%) \\
\vspace*{-8pt}\\
\hline 
\vspace*{-8pt}\\
Statistical & 6.63 \\
Chiral extrapolation & 0.89 \\
Quark mass tuning & 2.18 \\
Discretization & 4.16 \\
Kinematic & 9.31 \\
Matching & 0.28 \\
\vspace*{-10pt}\\
\hline 
\vspace*{-8pt}\\
Total & 13.03
\end{tabular}
\end{ruledtabular}
\end{table}
The sources of uncertainty listed in Table \ref{tab:ffratioerrors} are:
\paragraph{Statistical.} Statistical uncertainties include the two- and 
three-point correlator fit errors and those associated with the lattice spacing 
determination, $r_1$ and $r_1/a$. These effects are 
the second largest source of uncertainty in our results, and are dominated by the smaller
statistics available in the $B_s\to K$ analysis.
\paragraph{Chiral extrapolation.} Includes the uncertainties arising from extrapolation in both valence and sea 
quark masses and from the $B_s\to D_s$ chiral logarithms in the chiral-continuum 
extrapolation, corresponding to the fit parameters $c_j^{(i)}$ in 
Eqs.~\eqref{eq:DjK} and \eqref{eq:DjDs}.
\paragraph{Quark mass tuning.} These uncertainties arise from tuning 
the light and strange quark masses at finite lattice spacing and
partial quenching effects.
\paragraph{Discretization.} These effects include the $(aE_{X_s}/\pi)^n$, $(a/r_1)^n$, 
and $(am_c)^n$ terms in the modified $z$-expansion, corresponding to the fit
parameters $e_j^{(i)}$, $f_j^{(i)}$ and $m_j^{(i)}$ in Eqs.~\eqref{eq:DjK} and \eqref{eq:DjDs}.
\paragraph{Kinematic.} Uncertainties that arise from the $z$-expansion 
coefficients, including the Blaschke factors. These effects are 
the dominant source of uncertainty in our results, and again predominantly arise 
from the $B_s\to K$ channel.
\paragraph{Matching.} The perturbative matching 
uncertainties stemming from the truncation of the expansion of NRQCD-HISQ effective 
currents in terms of QCD currents. These are the second largest source of uncertainty in 
the results for the individual channels, but the effects largely cancel in the ratio. This is further demonstrated
by tests (xi) and (xii)~of the previous section, in which changing the matching uncertainty from
2\% to 5\% has practically negligible effect on the fit, and in particular, the ratio at zero momentum transfer.

We propagate all uncertainties from the large momentum-transfer region, for 
which we have lattice results, to zero momentum transfer. We do not include the uncertainties associated with physical meson mass input errors and finite 
volume effects, which are both less than $0.01\%$, because they are negligible contributions to our error budget
estimates. Moreover, we neglect uncertainties from isospin breaking, electromagnetic 
effects, and charm-quark quenching effects in the gauge ensembles.

\begin{figure}
\centering
\caption{\label{fig:fferrsq2}Error budget estimates for the ratios of the form factors, 
$f_0^{(K)}/f_0^{(D_s)}(q^2)$ (upper panel) and $f_+^{(K)}/f_+^{(D_s)}(q^2)$ (lower panel), as a function of the 
momentum transfer, $q^2$.}
\includegraphics[width=0.44\textwidth,keepaspectratio]{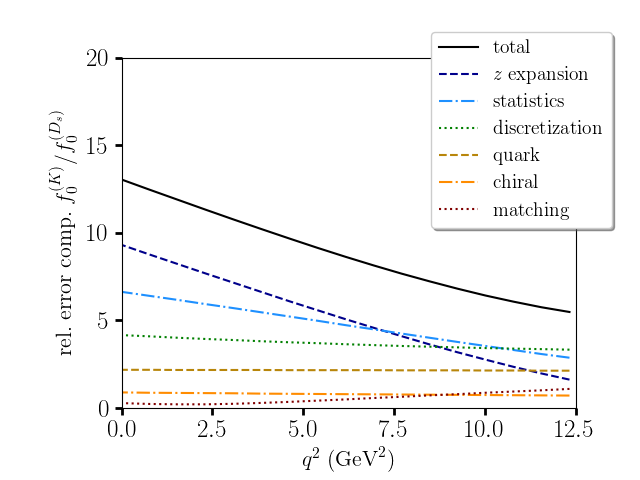}
\includegraphics[width=0.44\textwidth,keepaspectratio]{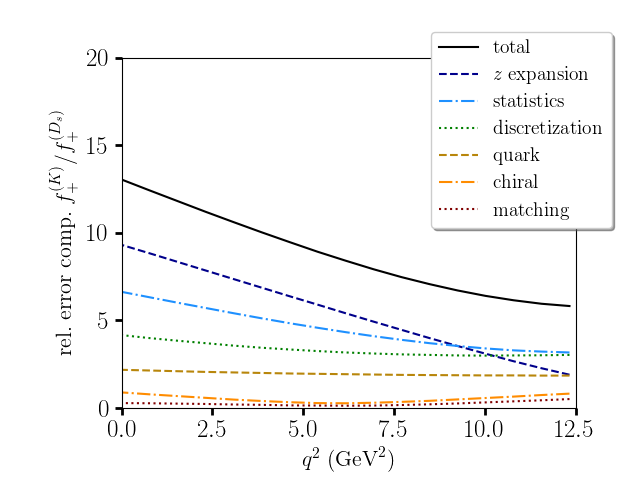}
\end{figure}

We plot our estimated error budges for the ratios of the form factors, $f_0(q^2)$ and $f_+(q^2)$, as a 
function of the momentum transfer, $q^2$, in Fig.~\ref{fig:fferrsq2}.

\subsection{Semileptonic decay phenomenology}

The experimental measurements of the ratio
\begin{equation}
R(D) = \frac{{\cal B}(B\to D\tau \nu)}{{\cal B}(B\to D\ell \nu)},
\end{equation}
which measures the ratio of branching fraction of the semileptonic decay to the 
$\tau$ lepton to the branching fraction to an electron or muon 
(represented by $\ell$), are currently in tension with the standard model result.
The global experimental average is 
\cite{Lees:2012xj,Lees:2013uzd,Huschle:2015rga,hfag:2016rds}
\begin{equation}
R(D)_{\mathrm{exp.}} = 0.391(41)_{\mathrm{stat.}}(28)_{\mathrm{sys.}},
\end{equation}
whereas the standard model expectation, neglecting correlations between the calculations 
\cite{Kamenik:2008tj,Bailey:2012rr,Na:2015kha}, is
\begin{equation}
R(D)_{\mathrm{theor.}} = 0.299(7).
\end{equation}

We determine the corresponding ratio of the $R$-ratios for the semileptonic $B_s\to X_s \ell \nu$ decays,
\begin{equation}
\frac{R(K)}{R(D_s)} = 2.02(12), 
\end{equation}
which is in agreement with, but with slightly smaller errors than, the value of $R(K)/R(D_s)= 0.695(50)/0.314(6) = 2.21(16)$ obtained assuming uncorrelated uncertainties between the values given in \cite{Bouchard:2014ypa,Monahan:2017uby}.

Neglecting final state electromagnetic interactions, the full angular dependence of the differential decay rate for $B_s\to X_s\ell\nu$ is given in terms of the corresponding scalar and vector form factors by
\begin{align}
\frac{\mathrm{d}^2\Gamma(B_s\to X_s\ell \nu)}{\mathrm{d}q^2\mathrm{d}\cos\theta_\ell}  {} & =\frac{G_F^2|V_{xb}|^2}{128\pi^3M_{B_s}}\left(1-\frac{m_\ell^2}{q^2}\right)^2|\vec{p}_{X_s}| \nonumber \\
\times \bigg[4M_{B_s}^2\vec{p}_{X_s}^2{} & \left(\sin^2\theta_\ell+\frac{m_\ell^2}{q^2}\cos^2\theta_\ell\right)|f_+|^2 \nonumber \\
+ \frac{4m_\ell^2}{q^2}{} & \left(M_{B_s}^2-M_{X_s}^2\right)M_{B_s}|\vec{p}_{X_s}|\cos\theta_\ell f_0f_+\nonumber \\
{} &+ \frac{m_\ell^2}{q^2}\left(M_{B_s}^2-M_{X_s}^2\right)^2|f_0|^2\bigg]. \label{eq:dGdq2dcos}
\end{align}
Here $\theta_\ell$ is defined as the angle between the final state lepton and the $B_s$ meson, in the frame in which $\vec{p}_\ell + \vec{p}_\nu = \vec{0}$. Integrating over the angle $\theta_\ell$, one obtains the standard model differential decay rate,
\begin{align}
\gamma_\ell^{(X_s)} {} &= \frac{\mathrm{d}^2\Gamma(B_s\to X_s\ell \nu)}{\mathrm{d}q^2 }  \nonumber \\
{} &=  \frac{G_F^2|V_{xb}|^2}{24\pi^3M_{B_s}}\left(1-\frac{m_\ell^2}{q^2}\right)^2|\vec{p}_{X_s}| \nonumber \\
{} & \quad \times \bigg[\left(1+\frac{m_\ell^2}{2q^2}\right)M_{B_s}^2\vec{p}_{X_s}^2|f_+|^2 \nonumber \\
{} & \qquad + \frac{3m_\ell^2}{8q^2}\left(M_{B_s}^2-M_{X_s}^2\right)^2|f_0|^2\bigg].
\end{align}

\begin{figure}
\centering
\caption{\label{fig:dGratioq2}Ratio of the differential decay rates, $\gamma_\ell^{(K)}/\gamma_\ell^{(D_s)}$, divided by $|V_{ub}/V_{cb}|^2$, as a function of the 
momentum transfer, $q^2$.}
\includegraphics[width=0.48\textwidth,keepaspectratio]{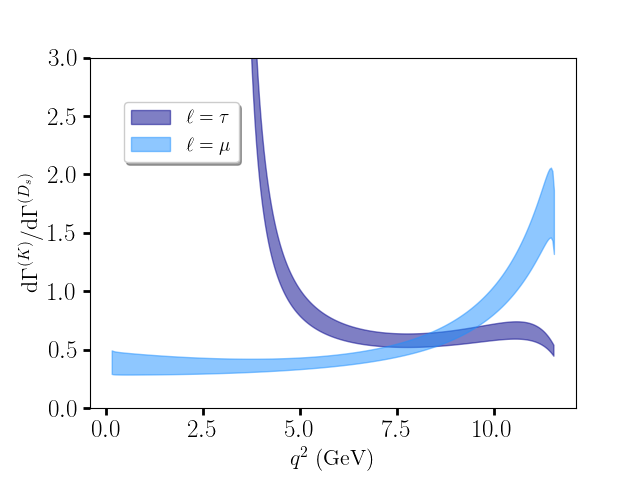}
\end{figure}

In Fig.~\ref{fig:dGratioq2} we plot the ratio of the differential decay rates, $\gamma_\ell^{(K)}/\gamma_\ell^{(D_s)}$,
as a function of the momentum transfer, for the semileptonic decays to muons ($\ell = \mu$) and to tau leptons ($\ell = \tau$).

\begin{figure}
\centering
\caption{\label{fig:dGVxb}Ratio of the differential decay rates, $\gamma_\ell^{(K)}/\gamma_\ell^{(D_s)}$, using inclusive and exclusive world average results for $|V_{ub}/V_{cb}|$, as a function of the 
momentum transfer, $q^2$. The upper panel shows the decay rates for $\ell = \tau$, and the lower panel $\ell = \mu$.}
\includegraphics[width=0.48\textwidth,keepaspectratio]{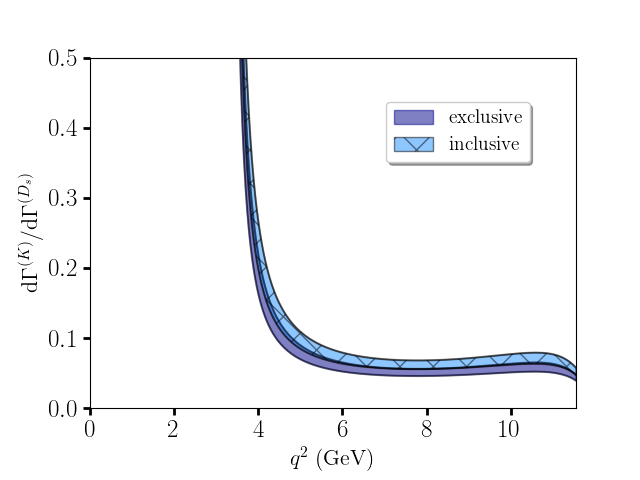}\\
\includegraphics[width=0.48\textwidth,keepaspectratio]{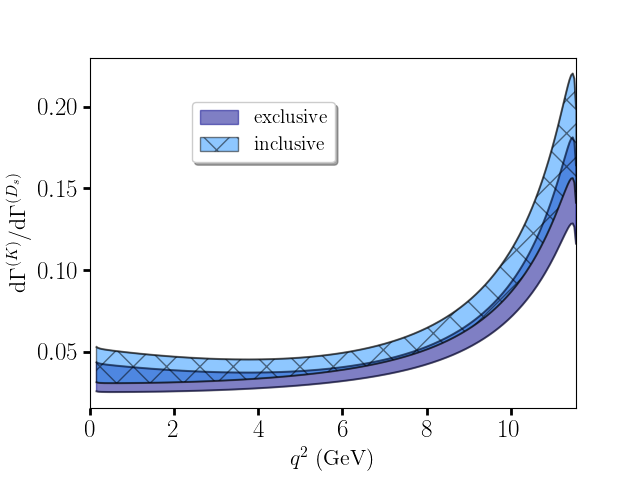}
\end{figure}

We combine our results for these decay rate ratios with the experimental world average results for $|V_{ub}/V_{cb}|$ \cite{pdg18}, using both inclusive and exclusive determinations,
\begin{align}
\mathrm{exclusive}\; \left|V_{ub}/V_{cb}\right| = {} & 0.088(6), \label{eq:vubvcbexc}\\ 
\mathrm{inclusive}\; \left|V_{ub}/V_{cb}\right| = {} & 0.107(7), \label{eq:vubvcbinc} 
\end{align}
and plot the results in Fig.~\ref{fig:dGVxb}. The LHCb Collaboration has measured this ratio to be $|V_{ub}/V_{cb}|=0.083(6)$, updated in \cite{pdg18} to $|V_{ub}/V_{cb}|=0.080(6)$, from the ratio of the baryonic semileptonic decays $\Lambda_b \to p^+\mu^-\overline{\nu}$ and $\Lambda_b \to \Lambda_c^+\mu^-\overline{\nu}$ \cite{Aaij:2015bfa}. This result is sufficiently close to the world average given in Eqs.~\eqref{eq:vubvcbexc} that we do not include it in Fig.~\ref{fig:dGVxb}. A correlated average, $|V_{ub}/V_{cb}|=0.092(8)$, of both inclusive and exclusive results is given in \cite{pdg18}, which also includes the experimental result from baryonic decays, but the large discrepancy between the inclusive and exclusive determinations suggests that this average should be treated with caution.

Defining the partially integrated ratio
\begin{equation}
\zeta^{(X_s)}_{\ell} = \frac{1}{|V_{xb}|^2}\int_{m_\ell^2}^{q^2_{\mathrm{max}}}\frac{\mathrm{d}\Gamma(B_s\to X_s\ell\nu)}{\mathrm{d}q^2}\mathrm{d}q^2,
\end{equation}
where $q^2_{\mathrm{max}} = (M_{B_s}- M_{X_s})^2$, we integrate our results numerically to obtain
\begin{align}
\frac{\zeta_\mu^{(K)}}{\zeta_\mu^{(D_s)}} = {} & 0.85(13), \\
\frac{\zeta_\tau^{(K)}}{\zeta_\tau^{(D_s)}} = {} & 1.72(19).
\end{align}

Asymmetries in the differential decay rate can be defined from the angular distribution, Eq.~\eqref{eq:dGdq2dcos}. The forward-backward asymmetry is given
by
\begin{align}
A_\ell^{(X_s)}(q^2) = {} & \left[\int_0^1 - \int_{-1}^0\right]\mathrm{d}\cos\theta_\ell \frac{\mathrm{d}^2\Gamma}{\mathrm{d}q^2\mathrm{d}\cos\theta_\ell} \nonumber \\
= {} & \frac{G_F^2|V_{xb}|^2}{32\pi^3M_{B_s}}\left(1-\frac{m_\ell^2}{q^2}\right)^2|\vec{p}_{X_s}|^2\frac{m_\ell^2}{q^2} \nonumber \\
{} & \qquad \times \left(M_{B_s}^2-M_{X_s}^2\right)f_0f_+,
\end{align}
and the polarization asymmetry by
\begin{equation}
P_\ell^{(X_s)}(q^2) = \frac{\mathrm{d}\Gamma(\mathrm{LH})}{\mathrm{d}q^2 }- \frac{\mathrm{d}\Gamma(\mathrm{RH})}{\mathrm{d}q^2 },
\end{equation}
where the differential decay rates to left-handed (LH) and right-handed (RH) final state leptons are given by
\begin{align}
\frac{\mathrm{d}\Gamma(\mathrm{LH})}{\mathrm{d}q^2 } = {} & \frac{G_F^2|V_{xb}|^2|\vec{p}_{X_s}|^3}{24\pi^3 }\left(1-\frac{m_\ell^2}{q^2}\right)^2 f_+^2, \\
\frac{\mathrm{d}\Gamma(\mathrm{RH})}{\mathrm{d}q^2 } = {} & \frac{G_F^2|V_{xb}|^2|\vec{p}_{X_s}|^3}{24\pi^3 }\frac{m_\ell^2}{q^2}\left(1-\frac{m_\ell^2}{q^2}\right)^2 \nonumber \\
\times {}& \left[\frac{3}{8}\frac{(M_{B_s}^2-M_{X_s}^2)^2}{M_{B_s}^2}f_0^2+\frac{1}{2}|\vec{p}_{X_s}|^2f_+^2\right].
\end{align}
In the standard model, the production of right-handed final state leptons is helicity suppressed, and so this asymmetry offers a probe for helicity-violating interactions generated by new physics. 

\begin{figure}
\centering
\caption{\label{fig:asym}Ratio of the forward-backward asymmetries, $A_\tau^{(K)}/A_\tau^{(D_s)}$ (upper panel) and $A_\mu^{(K)}/A_\mu^{(D_s)}$ (lower panel), using inclusive and exclusive world average results for $|V_{ub}/V_{cb}|$, as a function of the 
momentum transfer, $q^2$.}
\includegraphics[width=0.48\textwidth,keepaspectratio]{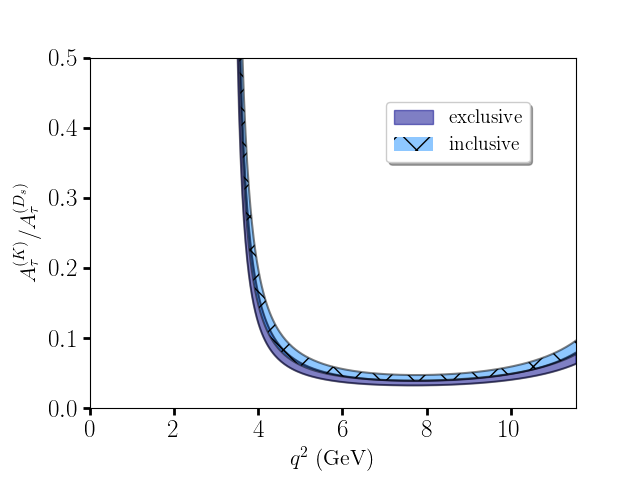}\\
\includegraphics[width=0.48\textwidth,keepaspectratio]{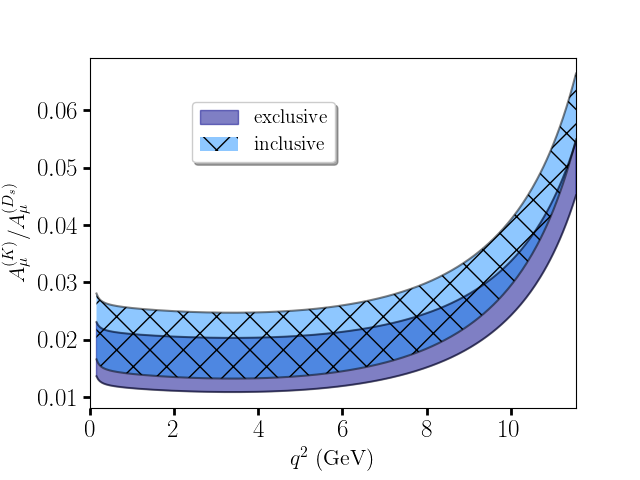}
\end{figure}

\begin{figure}
\centering
\caption{\label{fig:psym}Ratio of the polarization asymmetries, $P_\tau^{(K)}/P_\tau^{(D_s)}$ (upper panel) and $P_\mu^{(K)}/P_\mu^{(D_s)}$ (lower panel), using inclusive and exclusive world average results for $|V_{ub}/V_{cb}|$, as a function of the 
momentum transfer, $q^2$.}
\includegraphics[width=0.48\textwidth,keepaspectratio]{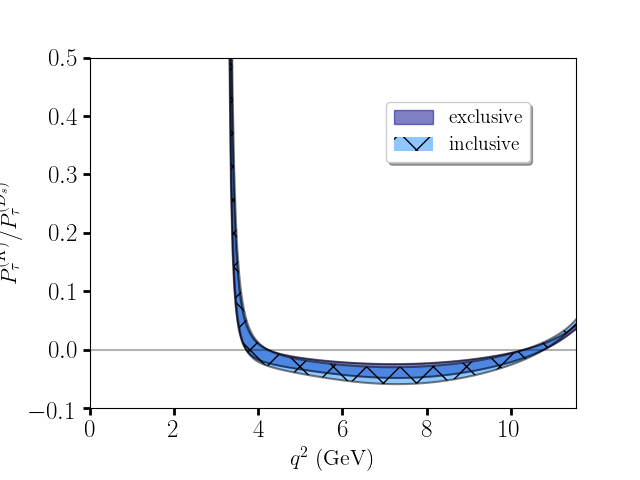}\\
\includegraphics[width=0.48\textwidth,keepaspectratio]{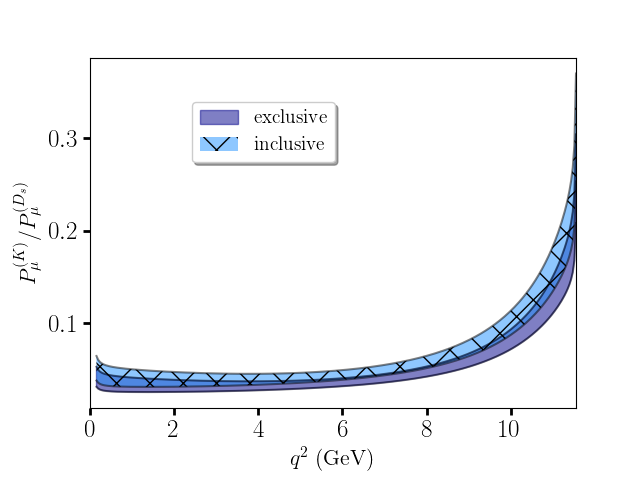}
\end{figure}

In Figs.~\ref{fig:asym} and \ref{fig:psym}, we plot the ratios of the forward-backward and polarization asymmetries, respectively, for the $B_s\to K\ell\nu$ and $B_s\to D_s\ell \nu$ decays. We plot the asymmetry ratios using both inclusive and exclusive values of $|V_{ub}/V_{cb}|$. Integrating over $q^2$, and multiplying by the appropriate combination of CKM matrix elements to define the QCD contribution, we find
\begin{align}
\frac{|V_{cb}|^2}{|V_{ub}|^2}\frac{\int_{m_\ell^2}^{q^2_{\mathrm{max}}}A^{(K)}_\mu \mathrm{d}q^2}{\int_{m_\mu^2}^{q^2_{\mathrm{max}}}A^{(D_s)}_\mu \mathrm{d}q^2} = {} & 0.399(85),\\
\frac{|V_{cb}|^2}{|V_{ub}|^2}\frac{\int_{m_\ell^2}^{q^2_{\mathrm{max}}}A^{(K)}_\tau \mathrm{d}q^2}{\int_{m_\tau^2}^{q^2_{\mathrm{max}}}A^{(D_s)}_\tau \mathrm{d}q^2} = {} & 1.38(15),\\
\frac{|V_{cb}|^2}{|V_{ub}|^2}\frac{\int_{m_\ell^2}^{q^2_{\mathrm{max}}}P^{(K)}_\mu \mathrm{d}q^2}{\int_{m_\mu^2}^{q^2_{\mathrm{max}}}P^{(D_s)}_\mu \mathrm{d}q^2} = {} & 0.87(13),\\
\frac{|V_{cb}|^2}{|V_{ub}|^2}\frac{\int_{m_\ell^2}^{q^2_{\mathrm{max}}}P^{(K)}_\tau \mathrm{d}q^2}{\int_{m_\tau^2}^{q^2_{\mathrm{max}}}P^{(D_s)}_\tau \mathrm{d}q^2} = {} & -0.42(22).
\end{align}

Normalizing these asymmetry ratios by the corresponding differential decay rate ratio removes the ambiguity arising from $|V_{ub}/V_{cb}|$,
\begin{align}
\overline{A}^{(X_s)}_\ell = {} & \frac{\int_{m_\ell^2}^{q^2_{\mathrm{max}}} A^{(X_s)}_\ell \mathrm{d}q^2}{\int_{m_\ell^2}^{q^2_{\mathrm{max}}}(\mathrm{d}\Gamma/\mathrm{d}q^2)\mathrm{d}q^2 },\\
\overline{P}^{(X_s)}_\ell = {} & \frac{\int_{m_\ell^2}^{q^2_{\mathrm{max}}} P^{(X_s)}_\ell \mathrm{d}q^2}{\int_{m_\ell^2}^{q^2_{\mathrm{max}}}(\mathrm{d}\Gamma/\mathrm{d}q^2)\mathrm{d}q^2 }.
\end{align}
We integrate over the momentum transfer numerically to find
\begin{align}
\frac{\overline{A}^{(K)}_\mu}{\overline{A}^{(D_s)}_\mu} = {} & 0.470(41),\qquad 
\frac{\overline{P}^{(K)}_\mu}{\overline{P}^{(D_s)}_\mu} = 1.0193(17)\\
\frac{\overline{A}^{(K)}_\tau}{\overline{A}^{(D_s)}_\tau} = {} & 0.804(15), \qquad
\frac{\overline{P}^{(K)}_\tau}{\overline{P}^{(D_s)}_\tau} = -0.25(11),
\end{align}
where the smaller relative uncertainties compared to the asymmetries themselves demonstrates that most of the hadronic uncertainties have canceled in these normalized results.

\section{\label{sec:summary}Summary}

We have presented a study of the ratio of the scalar and vector form 
factors for the $B_s \to 
X_s\ell\nu$ semileptonic decays, where $X_s$ is a $K$ or $D_s$ meson, over the full kinematic range of 
momentum transfer. These ratios combine correlator data results determined in \cite{Bouchard:2014ypa} for 
the $B_s \rightarrow K$ decay and in \cite{Monahan:2017uby} for the
$B_s\rightarrow D_s$ decay. Our simultaneous, correlated chiral-continuum kinematic extrapolation reduces the uncertainty
in the form factor ratio and, in particular, largely removes the uncertainty
arising from the perturbative matching procedure.

In addition to the form factor ratios, we predict $R(K)/R(D_s)$, where $R(X_s)$ is the ratio of the branching fractions 
of the corresponding semileptonic $B_s$ decay to tau and to electrons and muons. We determine the ratio of the 
differential decay rates for the two decay channels, as well as the ratio of the forward-backward and polarization
asymmetries.

The LHC is scheduled to significantly improve the statistical 
uncertainties in experimental measurements of $B_s$ decays with more data over 
the next decade. In particular, experimental data on the ratio of the $B_s \rightarrow K\ell\nu$ and
$B_s\rightarrow D_s \ell \nu$ decays, when combined with our form factor results, 
will provide a new determination of $|V_{ub}/V_{cb}|$.

\begin{acknowledgments}
Numerical simulations were carried out on facilities
of the USQCD Collaboration funded by the Office of Science of the Department of Energy and 
at the Ohio Supercomputer Center. Parts of this work were supported by the 
National Science Foundation. C.J.M. was supported in part
by the U.S.~Department of Energy through Grant No.~DE-FG02-00ER41132 and J.S. in part by the U.S.~Department of Energy through Grant No.~DE-SC0011726. We thank the MILC Collaboration for use of their gauge configurations.
\end{acknowledgments}

\appendix

\section{\label{app:twopt}Two-point fit results}

Here we reproduce the two-point fit results of \cite{Bouchard:2014ypa} in Table \ref{tab:k2pt} for the $K$ meson
and for the $D_s$ meson \cite{Monahan:2017uby} in Table \ref{tab:ds2pt}.
\begin{table}
\caption{\label{tab:k2pt}
Fit results for the ground state energies of the $K$ meson at each spatial 
momentum $\vec{p}_K$. Data reproduced from Table V of \cite{Bouchard:2014ypa}.}
\begin{ruledtabular}
\begin{tabular}{ccccc}
Set & $a M_K$   & $a E_K(1,0,0)$ & $a E_K(1,1,0)$ & $  
a E_K(1,1,1)$  \\
\vspace*{-10pt}\\
\hline 
\vspace*{-6pt}\\
C1 & 0.31211(15) & 0.40657(58) & 0.48461(76) & 0.5511(16) \\
C2 & 0.32863(18) & 0.54506(85) & 0.5511(16) & 0.6261(75) \\
C3 & 0.35717(22) & 0.47521(85) & 0.5723(11) & 0.6524(30) \\
F1 & 0.22865(11) & 0.32024(66) & 0.39229(86) & 0.4515(25) \\
F2 & 0.24577(13) & 0.33322(52) & 0.40214(73) & 0.4623(14) \\
\end{tabular}
\end{ruledtabular}
\end{table}
\begin{table}
\caption{\label{tab:ds2pt}
Fit results for the ground state energies of the $D_s$ meson at each spatial 
momentum $\vec{p}_{D_s}$. Data reproduced from Table IV of \cite{Monahan:2017uby}.}
\begin{ruledtabular}
\begin{tabular}{ccccc}
Set & $a M_{D_s}$   & $a E_{D_s}(1,0,0)$ & $a E_{D_s}(1,1,0)$ & $  
a E_{D_s}(1,1,1)$  \\
\vspace*{-10pt}\\
\hline 
\vspace*{-6pt}\\
C1 & 1.18755(22) & 1.21517(34) & 1.24284(33) & 1.27013(39) \\
C2 & 1.20090(30) & 1.24013(56) & 1.27822(61) & 1.31543(97) \\
C3 & 1.19010(33) & 1.23026(53) & 1.26948(54) & 1.30755(79) \\
F1 & 0.84674(12) & 0.87559(19) & 0.90373(20) & 0.93096(26) \\
F2 & 0.84415(14) & 0.87348(25) & 0.90145(25) & 0.92869(33) \\
\end{tabular}
\end{ruledtabular}
\end{table}

\section{\label{sec:ffdetails}Reconstructing form factors}

In this Appendix we provide our fit results for the coefficients of the $z$-expansion for the $B_s \to K \ell \nu$ decay in Table \ref{tab:BsKzexp}, for $B_s \to D_s \ell \nu$ in Table \ref{tab:BsDszexp}, and for the correlated fit to both decays in Table \ref{tab:BsKDszexp}. We also 
tabulate our choice of priors for the chiral-continuum extrapolation for the 
$B_s\to K\ell\nu$ decay in Tables \ref{tab:BsKzfit} and \ref{tab:KII}, for the
$B_s\to D_s\ell\nu$ decay in Tables \ref{tab:BsDszfit} and \ref{tab:DsII}, and for priors common to both channels in \ref{tab:grp2}, and \ref{tab:grp3}.
\begin{table*}
\caption{\label{tab:BsKzexp}
Coefficients of $z$-expansion and the corresponding Blaschke factors for the $B_s\to K\ell \nu$ decay.}
\begin{ruledtabular}
\begin{tabular}{cccccccc}
$a_1^{(0)}$ & $a_2^{(0)}$ & $a_3^{(0)}$ & $P_0$ & $a_0^{(+)}$ & $a_1^{(+)}$ 
& 
$a_2^{(+)}$ & $P_+$ \\
\vspace*{-8pt}\\
\hline
\vspace*{-8pt}\\
0.336(88) & 1.23(70) & 2.1(2.6) & 5.6793(10) & 0.301(18)  & -0.48(12) & 2.39(86) & 5.32450(27) \\ 
\end{tabular}
\end{ruledtabular}
\end{table*}

\begin{table*}
\caption{\label{tab:BsDszexp}
Coefficients of $z$-expansion and the corresponding Blaschke factors, for the $B_s\to D_s\ell \nu$ decay.}
\begin{ruledtabular}
\begin{tabular}{ccccccccc}
$a_0^{(0)}$ & $a_1^{(0)}$ & $a_2^{(0)}$ & $a_3^{(0)}$ & $P_0$ & $a_0^{(+)}$ & $a_1^{(+)}$ 
& 
$a_2^{(+)}$ & $P_+$ \\
\vspace*{-8pt}\\
\hline
\vspace*{-8pt}\\
0.673(39)  & -0.02(34) & 1.4(2.8) & -0.1(3.0) & 6.41(10) & 0.773(37) & -3.01(56) & 
-0.01(2.95) & 6.3300(90) \\ 
\end{tabular}
\end{ruledtabular}
\end{table*}

\begin{table*}
\caption{\label{tab:BsKDszexp}
Covariance matrix for the coefficients of $z$-expansion and the corresponding Blaschke factors for the simultaneous fit to the $B_s\to K\ell \nu$ and $B_s\to D_s\ell \nu$ decays. The rows 
correspond to the columns, moving from top to bottom and left to right, 
respectively.
}
\begin{tabular}{cccccc}
\hline\vspace{-8pt}\\
\hline \vspace*{-8pt}\\
$a_1^{(0),K}$ & $a_2^{(0),K}$ & $a_3^{(0),K}$ & $P_0^{(K)}$ 
& $a_0^{(+),K}$ & $a_1^{(+),K}$ \\
\vspace*{-8pt}\\
\hline
\vspace*{-8pt}\\
7.81655746$\times 10^{-3}$ & 5.11931999$\times 10^{-2}$ & 1.26040746$\times 10^{-1}$ & -3.95599616$\times 10^{-7}$
 &  6.67729571$\times 10^{-4}$ & 7.88936302$\times 10^{-3}$ \\
& 4.94505240$\times 10^{-1}$ & 1.62865239 & 2.22974369$\times 10^{-6}$ &  3.58512534$\times 10^{-3}$ & 6.75709862$\times 10^{-2}$ \\
& & 6.51816994 & -4.88348307$\times 10^{-8}$ & 9.03252850$\times 10^{-3}$ & 1.99167048$\times 10^{-1}$ \\
& & & 9.99995307$\times 10^{-7}$ & -1.81816269$\times 10^{-9}$ & 1.55891061$\times 10^{-7}$ \\
& & & &  3.09228616$\times 10^{-4}$ & -5.88646696$\times 10^{-5}$ \\
& & & & & 1.46893824$\times 10^{-2}$ \\
\\
\end{tabular}
\vspace*{\baselineskip}
\begin{tabular}{cccccc}
$a_2^{(+),K}$ & $P_+^{(K)}$ 
& $a_0^{(0),D_s}$ & $a_1^{(0),D_s}$ & $a_2^{(0),D_s}$ & $a_3^{(0),D_s}$ \\
\vspace*{-8pt}\\
\hline
\vspace*{-8pt}\\
5.54055868$\times 10^{-2}$ & 5.22263419$\times 10^{-9}$
 &  4.89761879$\times 10^{-5}$ & 1.47978430$\times 10^{-3}$ & 1.61294090$\times 10^{-4}$ & -1.50864482$\times 10^{-5}$
\\
5.20212224$\times 10^{-1}$ & 4.60220124$\times 10^{-8}$
 &  4.23550639$\times 10^{-4}$ & -1.12557927$\times 10^{-3}$ & -4.15916006$\times 10^{-4}$ & 6.86722615$\times 10^{-6}$
 \\
1.72576055 & 1.64613013$\times 10^{-7}$
 &  5.32746249$\times 10^{-4}$ & -8.00096682$\times 10^{-3}$ & -1.57760368$\times 10^{-3}$ & 6.07028861$\times 10^{-5}$
 \\
1.27709131$\times 10^{-6}$ & 4.34812507$\times 10^{-15}$
 & -2.93868039$\times 10^{-9}$ & 3.60812633$\times 10^{-8}$ & 3.12552274$\times 10^{-9}$ & -2.93053824$\times 10^{-10}$
 \\
5.57789886$\times 10^{-4}$ & 3.44350904$\times 10^{-9}$
 &  1.08803466$\times 10^{-4}$ & 7.14515361$\times 10^{-4}$ & 1.46191770$\times 10^{-4}$ & -9.57576314$\times 10^{-6}$
\\
6.49789179$\times 10^{-2}$ & -1.42002142$\times 10^{-7}$
 &  2.37456520$\times 10^{-4}$ & -7.74705909$\times 10^{-3}$ & -1.63296714$\times 10^{-3}$ & 9.27876845$\times 10^{-5}$
\\
7.40157233$\times 10^{-1}$ & 8.20182628$\times 10^{-7}$
 &  9.33127619$\times 10^{-4}$ & 3.38332719$\times 10^{-4}$ & -1.12948406$\times 10^{-5}$ & -1.17310027$\times 10^{-5}$
\\
& 5.28997606$\times 10^{-8}$
 & -4.00252884$\times 10^{-11}$ & 1.55683903$\times 10^{-10}$ & 8.25859041$\times 10^{-11}$ & 4.62131689$\times 10^{-12}$
\\
& &  1.51331616$\times 10^{-3}$ & -1.32946477$\times 10^{-3}$ & -2.95921529$\times 10^{-3}$ & -1.18940865$\times 10^{-4}$
\\
& & & 1.14391084$\times 10^{-1}$ & 3.77594136$\times 10^{-1}$ & -1.47064962$\times 10^{-2}$
\\
& & & & 8.04802477 & 6.00685427$\times 10^{-2}$
 \\
& & & & & 8.99580234 
\end{tabular}
\vspace*{\baselineskip}
\begin{tabular}{ccccc}
$P_0^{(D_s)}$ 
& $a_0^{(+),D_s}$ & $a_1^{(+),D_s}$ & $a_2^{(+),D_s}$ & $P_+^{(D_s)}$ \\
\vspace*{-8pt}\\
\hline
\vspace*{-8pt}\\
2.48190307$\times 10^{-6}$ & 1.25952168$\times 10^{-4}$ & -1.00202940$\times 10^{-3}$ & 3.13648146$\times 10^{-5}$
 & -1.42966100$\times 10^{-8}$ \\
1.66495291$\times 10^{-6}$ & 4.16420952$\times 10^{-4}$ & -8.93653944$\times 10^{-4}$ & 4.32425257$\times 10^{-4}$
 & -3.15809640$\times 10^{-8}$ \\
-3.14364934$\times 10^{-6}$ & 2.30951064$\times 10^{-4}$ & 1.62406281$\times 10^{-3}$ & 1.00576304$\times 10^{-3}$
 &  8.26930192$\times 10^{-10}$ \\
5.51018709$\times 10^{-11}$ & -1.49607346$\times 10^{-9}$ & -1.05105378$\times 10^{-8}$ & 3.28268609$\times 10^{-10}$
 & -4.41710197$\times 10^{-14}$ \\
7.04107924$\times 10^{-7}$ & 1.61718771$\times 10^{-4}$ & -9.47821843$\times 10^{-4}$ & -7.78344712$\times 10^{-6}$
 & -1.33434640$\times 10^{-8}$ \\
-6.39749633$\times 10^{-6}$ & -1.59677031$\times 10^{-4}$ & 4.95080220$\times 10^{-3}$ & 5.36015327$\times 10^{-4}$
 &  2.94137887$\times 10^{-8}$ \\
2.60283535$\times 10^{-6}$ & 1.06965630$\times 10^{-3}$ & -3.87458330$\times 10^{-3}$ & 3.61587037$\times 10^{-4}$
 & -8.16004467$\times 10^{-8}$ \\
-1.19874155$\times 10^{-12}$ & -3.95882072$\times 10^{-11}$ & 2.69588869$\times 10^{-10}$ & -1.34953976$\times 10^{-10}$
 &  2.49776863$\times 10^{-15}$ \\
3.86973615$\times 10^{-4}$ & 1.25442551$\times 10^{-3}$ & 7.19766977$\times 10^{-3}$ & 7.34363847$\times 10^{-3}$
 & -6.23834522$\times 10^{-7}$ \\
-1.51873367$\times 10^{-2}$ & 1.80319837$\times 10^{-3}$ & 2.26955835$\times 10^{-2}$ & 2.65518223$\times 10^{-2}$
 & -2.61624148$\times 10^{-6}$ \\
4.53224736$\times 10^{-3}$ & 1.19829415$\times 10^{-2}$ & 1.18533968$\times 10^{-1}$ & 2.29348564$\times 10^{-1}$
 & -2.65313186$\times 10^{-5}$ \\
-2.72916676$\times 10^{-4}$ & -1.76329307$\times 10^{-4}$ & -1.96104068$\times 10^{-3}$ & -8.21918389$\times 10^{-3}$
 &  1.15675448$\times 10^{-6}$ \\
9.95331216$\times 10^{-3}$ & -5.12869129$\times 10^{-5}$ & -5.75838181$\times 10^{-4}$ & -9.37738726$\times 10^{-4}$
 &  1.10417128$\times 10^{-7}$\\
& 1.37380763$\times 10^{-3}$ & -1.31877655$\times 10^{-3}$ & -8.10703811$\times 10^{-3}$
 &  4.47444315$\times 10^{-6}$ \\
& & 3.21831236$\times 10^{-1}$ & 2.71750438$\times 10^{-1}$
 & -1.70915346$\times 10^{-4}$ \\
& & & 8.72142922 &  4.09895485$\times 10^{-5}$ \\
& & & &  8.10107530$\times 10^{-5}$\\
\hline\vspace{-8pt}\\
\hline
\end{tabular}

\end{table*}

\begin{table}
\caption{\label{tab:BsKzfit}
Group I priors and fit results for the parameters in the 
modified $z$-expansion for the 
$B_s\to K\ell \nu$ decay. Note that these parameters are fit simultaneously with
those of Table \ref{tab:BsDszfit}, but displayed separately for clarity.
}
\begin{ruledtabular}
\begin{tabular}{ccccc}
& Prior $[f_0]$ & Fit result $[f_0]$ & Prior $[f_+]$ & Fit result $[f_+]$ 
\\
\vspace*{-10pt}\\
\hline 
\vspace*{-8pt}\\
$a_1$ & 0.0(3.0) & 0.336(88) & 0.0(5.0) & 0.301(43) \\
$a_2$ & 0.0(3.0) & 1.23(70) & 0.0(5.0) & -0.48(23) \\ 
$a_3$ & 0.0(3.0) & 2.1(2.6) & 0.0(5.0) & 2.39(86) \\ 
$c_1^{(1)}$ & 0.0(1.0) & -0.17(48) & 0.0(1.0) &  0.222(89) \\
$c_2^{(1)}$ & 0.0(1.0) & 0.34(72) & 0.0(1.0) & 0.52(48) \\ 
$c_3^{(1)}$ & 0.0(1.0) &  0.002(976) & 0.0(1.0) & -0.11(65) \\ 
$d_1^{(1)}$ & 0.00(30) &  -0.08(30) & 0.00(30) & 0.03(26) \\
$d_2^{(1)}$ & 0.00(30) & 0.02(30) & 0.00(30) & 0.02(30) \\
$d_3^{(1)}$ & 0.00(30) & 0.002(300) & 0.00(30) & 0.02(30) \\ 
$d_1^{(2)}$ & 0.0(1.0) & 0.3(1.0) & 0.00(30) & 0.04(97) \\
$d_2^{(2)}$ & 0.0(1.0) & -0.2(1.0) & 0.00(30) & 0.007(1.0) \\
$d_3^{(2)}$ & 0.0(1.0) & 0.04(1.0) & 0.00(30) & 0.004(1.0) \\
$e_1^{(1)}$ & 0.00(30) & 0.0007(0.3) & 0.00(30) & 0.013(28) \\
$e_2^{(1)}$ & 0.00(30) & 0.006(0.3) & 0.00(30) & 0.0007(0.3) \\
$e_3^{(1)}$ & 0.00(30) & -0.002(0.3) & 0.00(30) & -0.003(0.3) \\
$e_1^{(2)}$ & 0.0(1.0) & 0.006(1.0) & 0.0(1.0) & 0.01(30) \\
$e_2^{(2)}$ & 0.0(1.0) & -0.001(1.0) & 0.0(1.0) & 0.0005(1.0) \\
$e_3^{(2)}$ & 0.0(1.0) & -0.0001(1.0) & 0.0(1.0) & 3$\times10^{-5}$(1.0) \\
$f_1^{(1)}$ & 0.00(30) & -0.20(27) & 0.00(30) & 0.24(19) \\
$f_2^{(1)}$ & 0.00(30) & 0.14(29) & 0.00(30) & -0.08(29) \\
$f_3^{(1)}$ & 0.00(30) & -0.03(30) & 0.00(30) & -0.03(30) \\
$f_1^{(2)}$ & 0.0(1.0) & -0.47(94) & 0.0(1.0) & 0.28(83) \\
$f_2^{(2)}$ & 0.0(1.0) & 0.33(98) & 0.0(1.0) & -0.13(99) \\
$f_3^{(2)}$ & 0.0(1.0) & -0.08(1.0) & 0.0(1.0) & -0.08(1.0) \\
$l_1^{(1,1)}$ & 0.0(1.0) & -0.21(98) & 0.0(1.0) & 0.09(99) \\
$l_2^{(1,1)}$ & 0.0(1.0) & -0.06(1.0) & 0.0(1.0) & 0.03(1.0) \\
$l_3^{(1,1)}$ & 0.0(1.0) & -0.0005(1.0) & 0.0(1.0) & 0.002(1.0) \\
$l_1^{(1,2)}$ & 0.0(1.0) & -0.07(1.0) & 0.0(1.0) & 0.06(1.0) \\
$l_2^{(1,2)}$ & 0.0(1.0) & -0.02(1.0) & 0.0(1.0) & 0.002(1.0) \\
$l_3^{(1,2)}$ & 0.0(1.0) & -0.0006(1.0) & 0.0(1.0) & -0.0003(1.0) \\
$l_1^{(2,1)}$ & 0.0(1.0) & -0.06(1.0) & 0.0(1.0) & -0.02(1.0) \\
$l_2^{(2,1)}$ & 0.0(1.0) & -0.003(1.0) & 0.0(1.0) & 0.009(1.0) \\
$l_3^{(2,1)}$ & 0.0(1.0) & 0.0007(1.0) & 0.0(1.0) & 0.003(1.0) \\
$l_1^{(2,2)}$ & 0.0(1.0) & -0.03(1.0) & 0.0(1.0) & -0.0003(1.0) \\
$l_2^{(2,2)}$ & 0.0(1.0) & -0.007(1.0) & 0.0(1.0) & 0.001(1.0) \\
$l_3^{(2,2)}$ & 0.0(1.0) & -0.0002(1.0) & 0.0(1.0) & 0.0004(1.0) \\
$h_1^{(1,1)}$ & 0.0(1.0) & -0.21(98) & 0.0(1.0) & -0.49(61) \\
$h_2^{(1,1)}$ & 0.0(1.0) & -0.06(1.0) & 0.0(1.0) & 0.2(1.0) \\
$h_3^{(1,1)}$ & 0.0(1.0) & -0.0005(1.0) & 0.0(1.0) & 0.03(1.0) \\
$h_1^{(1,2)}$ & 0.0(1.0) & -0.07(1.0) & 0.0(1.0) & -0.04(97) \\
$h_2^{(1,2)}$ & 0.0(1.0) & -0.02(1.0) & 0.0(1.0) & 0.03(1.0) \\
$h_3^{(1,2)}$ & 0.0(1.0) & -0.0006(1.0) & 0.0(1.0) & 0.004(1.0) \\
$h_1^{(2,1)}$ & 0.0(1.0) & -0.06(1.0) & 0.0(1.0) & -0.11(1.0) \\
$h_2^{(2,1)}$ & 0.0(1.0) & -0.003(1.0) & 0.0(1.0) & 0.03(1.0) \\
$h_3^{(2,1)}$ & 0.0(1.0) & 0.0007(1.0) & 0.0(1.0) & 0.01(1.0) \\
$h_1^{(2,2)}$ & 0.0(1.0) & -0.03(1.0) & 0.0(1.0) & -0.04(1.0) \\
$h_2^{(2,2)}$ & 0.0(1.0) & -0.007(1.0) & 0.0(1.0) & 0.01(1.0) \\
$h_3^{(2,2)}$ & 0.0(1.0) & -0.0002(1.0) & 0.0(1.0) & 0.003(1.0) \\
\end{tabular}
\end{ruledtabular}
\end{table}

\begin{table}
\caption{\label{tab:BsDszfit}
Group I priors and fit results for the parameters in the 
modified $z$-expansion for the 
$B_s\to D_s\ell \nu$ decay. Note that these parameters are fit simultaneously with
those of Table \ref{tab:BsKzfit}, but displayed separately for clarity.
}
\begin{ruledtabular}
\begin{tabular}{ccccc}
& Prior $[f_0]$ & Fit result $[f_0]$ & Prior $[f_+]$ & Fit result $[f_+]$ 
\\
\vspace*{-10pt}\\
\hline 
\vspace*{-8pt}\\
$a_0$ & 0.0(3.0) & 0.673(39) & 0.0(5.0) & 0.773(37) \\ 
$a_1$ & 0.0(3.0) & -0.02(34) & 0.0(5.0) & -3.01(56) \\
$a_2$ & 0.0(3.0) & 1.4(2.8) & 0.0(5.0) & -0.01(2.95) \\ 
$a_3$ & 0.0(3.0) & -0.1(3.0) & - & - \\ 
$c_0^{(1)}$ & 0.0(1.0) &  0.087(15) & 0.0(1.0) &  0.188(69) \\
$c_1^{(1)}$ & 0.0(1.0) & -0.03(1.0) & 0.0(1.0) & 0.61(46) \\ 
$c_2^{(1)}$ & 0.0(1.0) & -0.09(1.0) & 0.0(1.0) & -0.0001(1.0) \\ 
$c_3^{(1)}$ & 0.0(1.0) & -0.0002(1.0) & - & - \\ 
$c_0^{(2)}$ & 0.00(30) & 0.006(31) & 0.00(30) & 0.165(67) \\
$c_1^{(2)}$ & 0.00(30) & 0.003(300) & 0.00(30) & 0.06(29) \\
$c_2^{(2)}$ & 0.00(30) & 0.005(30) & 0.00(30) & -7$\times10^{-6}$(0.3) \\
$c_3^{(2)}$ & 0.00(30) & 5$\times10^{-6}$(0.3) & - & - \\  
$d_0^{(1)}$ & 0.00(30) & -0.36(16) & 0.00(30) & -0.52(17) \\
$d_1^{(1)}$ & 0.00(30) & -0.0006(0.3) & 0.00(30) & -0.03(30) \\
$d_2^{(1)}$ & 0.00(30) & -0.0002(0.3) & 0.00(30) & 2$\times10^{-6}$(0.3) \\
$d_3^{(2)}$ & 0.00(30) & 3$\times10^{-6}$(0.3) & - & - \\
$d_0^{(2)}$ & 0.00(30) & 0.06(30) & 0.00(30) & 0.11(30) \\
$d_1^{(2)}$ & 0.00(30) & 7$\times10^{-5}$(0.3) & 0.00(30) & 0.01(30) \\
$d_2^{(2)}$ & 0.00(30) & 1$\times10^{-4}$(0.3) & 0.00(30) & 
-1$\times10^{-6}$(0.3) \\
$d_2^{(3)}$ & 0.00(30) & 2$\times10^{-7}$(0.3) & - & - \\
$e_0^{(1)}$ & 0.00(30) & 0.17(25) & 0.00(30) & 0.18(23) \\
$e_1^{(1)}$ & 0.00(30) & -0.0008(0.3) & 0.00(30) & -0.02(30) \\
$e_2^{(1)}$ & 0.00(30) & 0.0008(0.3) & 0.00(30) & 5$\times10^{-6}$(0.3) \\
$e_3^{(1)}$ & 0.00(30) & 1$\times10^{-5}$(0.3) & - & - \\
$e_0^{(2)}$ & 0.0(1.0) & 1.51(53) & 0.0(1.0) & 0.06(29) \\
$e_1^{(2)}$ & 0.0(1.0) & -0.002(1.0) & 0.0(1.0) & -0.001(1.0) \\
$e_2^{(2)}$ & 0.0(1.0) & -0.002(1.0) & 0.0(1.0) & 1$\times10^{-6}$(1.0) \\
$e_3^{(2)}$ & 0.0(1.0) & 9$\times10^{-6}$(1.0) & - & - \\
$m_0^{(1)}$ & 0.00(30) & -0.004(0.229) & 0.00(30) & 0.15(23) \\
$m_1^{(1)}$ & 0.00(30) & -0.0003(0.3) & 0.00(30) & -0.09(28) \\
$m_2^{(1)}$ & 0.00(30) & 0.008(0.3) & 0.00(30) & 2$\times10^{-5}$(0.3) \\
$m_3^{(1)}$ & 0.00(30) & 4$\times10^{-5}$(0.3) & - & - \\
$m_0^{(2)}$ & 0.0(1.0) & -0.49(40) & 0.0(1.0) & -0.34(42) \\
$m_1^{(2)}$ & 0.0(1.0) & -0.003(1.0) & 0.0(1.0) & -0.74(81) \\
$m_2^{(2)}$ & 0.0(1.0) & 0.03(1.0) & 0.0(1.0) & 0.0001(1.0) \\
$m_3^{(2)}$ & 0.0(1.0) & 0.0002(1.0) & - & - \\
\end{tabular}
\end{ruledtabular}
\end{table}

\begin{table}
\caption{\label{tab:KII}
Group II priors and fit results for the parameters in the 
modified $z$-expansion for the $B_s\to K\ell \nu$ decay.
}
\begin{ruledtabular}
\begin{tabular}{ccc}
Quantity & Prior  & Fit result
\\
\vspace*{-10pt}\\
\hline 
\vspace*{-8pt}\\
$aE_K(0,0,0)$ & 0.31195(14) & 0.31197(14) \\
& 0.32870(17) & 0.32865(17) \\ 
& 0.35744(21) & 0.35747(21) \\ 
& 0.22861(12) & 0.22862(12) \\
& 0.24566(13) & 0.24565(13) \\
$aE_K(1,0,0)$ & 0.40661(49) & 0.40662(48) \\
& 0.45434(73) & 0.45432(70) \\
& 0.47507(71) & 0.47566(69) \\
& 0.32020(61) & 0.31986(58) \\
& 0.33310(50) & 0.33293(49) \\
$aE_K(1,1,0)$ & 0.48408(63) & 0.48393(62) \\
& 0.5506(11) & 0.5511(11) \\
& 0.57218(80) & 0.57168(78) \\
& 0.39192(82) & 0.39240(79) \\
& 0.40184(72) & 0.40204(70) \\
$aE_K(1,1,1)$ &  0.5513(13) & 0.5511(13) \\
& 0.6273(35) & 0.6290(34) \\
& 0.6539(18) & 0.6534(17) \\
& 0.4528(16) & 0.4527(15) \\
& 0.4624(11) & 0.4624(11) \\
$M_+$ & 5.32450(27) & 5.32450(27) \\
$M_0$ & 5.6793(10) & 5.6793(10) 
\end{tabular}
\end{ruledtabular}
\end{table}

\begin{table}
\caption{\label{tab:DsII}
Group II priors and fit results for the parameters in the 
modified $z$-expansion for the $B_s\to D_s\ell \nu$ decay.
}
\begin{ruledtabular}
\begin{tabular}{ccc}
Quantity & Prior  & Fit result
\\
\vspace*{-10pt}\\
\hline 
\vspace*{-8pt}\\
$aE_{D_s}(0,0,0)$ & 1.18750(15) & 1.18749(15) \\
& 1.20126(21) & 1.20132(20) \\ 
& 1.19031(24) & 1.19020(24) \\ 
& 0.84674(12) & 0.84674(12) \\
& 0.84419(10) & 0.84421(10) \\
$aE_{D_s}(1,0,0)$ & 1.21497(19) & 1.21504(19) \\
& 1.24055(30) & 1.24080(28) \\
& 1.23055(35) & 1.23055(31) \\
& 0.87575(18) & 0.87574(18) \\
& 0.87353(16) & 0.87345(15) \\
$aE_{D_s}(1,1,0)$ & 1.24264(19) & 1.24274(19) \\
& 1.27942(29) & 1.27958(26) \\
& 1.26974(35) & 1.26941(32) \\
& 0.90393(18) & 0.90392(18) \\
& 0.90144(16) & 0.90148(15) \\
$aE_{D_s}(1,1,1)$ & 1.26988(22) & 1.26997(22) \\
& 1.31755(46) & 1.31737(39) \\
& 1.30768(48) & 1.30727(41) \\
& 0.93126(24) & 0.93123(24) \\
& 0.92873(20) & 0.92880(20) \\
$M_+$ & 6.3300(90) & 6.3300(90) \\
$M_0$ & 6.41(10) & 6.41(10) \\
\end{tabular}
\end{ruledtabular}
\end{table}

\begin{table}
\caption{\label{tab:grp2}
Group II priors and fit results for the parameters in the 
modified $z$-expansion, common to both $B_s\to X_s\ell \nu$ decay channels.
}
\begin{ruledtabular}
\begin{tabular}{ccc}
Quantity & Prior  & Fit result
\\
\vspace*{-10pt}\\
\hline 
\vspace*{-8pt}\\
$r_1/a$ & 2.6470(30) & 2.6463(28) \\
& 2.6180(30) & 2.6209(27) \\ 
& 2.6440(30) & 2.6423(29) \\
& 3.6990(30) & 3.6984(30) \\
& 3.7120(40) & 3.7127(40) \\
$aM_B$ & 3.18915(65) & 3.18921(64) \\ 
& 3.23184(88) & 3.23136(85) \\
& 3.21191(77) & 3.21221(76) \\ 
& 2.28109(52) & 2.28120(51) \\  
& 2.28101(44) & 2.28093(44) \\
$aM_{B_s}$ & 3.23019(25) & 3.23012(25) \\
& 3.26785(33) & 3.26792(33) \\ 
& 3.23585(38) & 3.23566(38) \\ 
& 2.30906(26) & 2.30899(25) \\ 
& 2.30122(16) & 2.30124(16) \\
$aM_\pi$ &  0.15990(20) &  0.15990(20) \\
& 0.21110(20) & 0.21110(20)  \\
& 0.29310(20) & 0.29310(20)  \\
&  0.13460(10) & 0.13460(10)  \\
& 0.18730(10) & 0.18730(10) \\
$aM_{\eta_s}$ & 0.41113(18) &  0.41113(18)  \\
&  0.41435(22) & 0.41433(22)  \\
&  0.41185(22) & 0.41186(22) \\
& 0.29416(12) &  0.29416(12)  \\
&  0.29311(18) & 0.29311(18)  \\
$aM_\pi^{\mathrm{MILC}}$ &  0.15971(20) &  0.15971(20)  \\
& 0.22447(17) & 0.22448(17) \\
& 0.31125(16) & 0.31124(16) \\
& 0.14789(18) & 0.14789(18) \\
& 0.20635(18) & 0.20636(18) \\
$aM_K^{\mathrm{MILC}}$ &  0.36530(29) & 0.36526(29)  \\
& 0.38331(24) & 0.38337(24)  \\
& 0.40984(21) & 0.40981(21)  \\
& 0.25318(19) &  0.25317(19)  \\
& 0.27217(21) & 0.27219(21)  \\
$1+m_\parallel$ & 1.000(40) & 1.001(40)   \\
$1+m_\perp$ & 1.000(40) & 1.000(40)
\end{tabular}
\end{ruledtabular}
\end{table}

\begin{table}
\caption{\label{tab:grp3}
Group III priors and fit results for the parameters in the 
modified $z$-expansion, common to both 
$B_s\to K\ell \nu$ and
$B_s\to D_s\ell \nu$ decays.
}
\begin{ruledtabular}
\begin{tabular}{ccc}
Quantity & Prior (GeV)  & Fit result (GeV)
\\
\vspace*{-10pt}\\
\hline 
\vspace*{-8pt}\\
$r_1$ & 0.3129(23) & 0.3130(23) \\
$M_\pi^{\mathrm{phys}}$ & 0.13497700(50) & 0.13497700(50) \\
$M_K^{\mathrm{phys}}$ & 0.495644(26) & 0.495644(26) \\
$M_\eta^{\mathrm{phys}}$ & 0.547862(17) & 0.547862(17) \\
$M_{\eta_s}^{\mathrm{phys}}$ & 0.6858(40) & 0.6857(40) \\
$M_{D_s}^{\mathrm{phys}}$ & 1.96828(10) & 1.96828(10) \\
$M_{B_s}^{\mathrm{phys}}$ & 5.36689(23) & 5.36689(23) \\
\end{tabular}
\end{ruledtabular}
\end{table}

\clearpage
\bibliographystyle{apsrev}
\bibliography{bstoxs.bib}

\begin{thebibliography}{68}
\expandafter\ifx\csname natexlab\endcsname\relax\def\natexlab#1{#1}\fi
\expandafter\ifx\csname bibnamefont\endcsname\relax
  \def\bibnamefont#1{#1}\fi
\expandafter\ifx\csname bibfnamefont\endcsname\relax
  \def\bibfnamefont#1{#1}\fi
\expandafter\ifx\csname citenamefont\endcsname\relax
  \def\citenamefont#1{#1}\fi
\expandafter\ifx\csname url\endcsname\relax
  \def\url#1{\texttt{#1}}\fi
\expandafter\ifx\csname urlprefix\endcsname\relax\def\urlprefix{URL }\fi
\providecommand{\bibinfo}[2]{#2}
\providecommand{\eprint}[2][]{\url{#2}}

\bibitem[{\citenamefont{Tanabashi et~al.}(2018)}]{pdg18}
\bibinfo{author}{\bibfnamefont{M.}~\bibnamefont{Tanabashi}}
  \bibnamefont{et~al.} (\bibinfo{collaboration}{Particle Data Group}),
  \bibinfo{journal}{Phys. Rev. D} \textbf{\bibinfo{volume}{98}},
  \bibinfo{pages}{030001} (\bibinfo{year}{2018}).

\bibitem[{\citenamefont{Amhis et~al.}(2017)}]{Amhis:2016xyh}
\bibinfo{author}{\bibfnamefont{Y.}~\bibnamefont{Amhis}} \bibnamefont{et~al.}
  (\bibinfo{collaboration}{HFLAV}), \bibinfo{journal}{Eur. Phys. J.}
  \textbf{\bibinfo{volume}{C77}}, \bibinfo{pages}{895} (\bibinfo{year}{2017}),
  \eprint{1612.07233}.

\bibitem[{\citenamefont{Aubert et~al.}(2006)}]{Aubert:2006ry}
\bibinfo{author}{\bibfnamefont{B.}~\bibnamefont{Aubert}} \bibnamefont{et~al.}
  (\bibinfo{collaboration}{BaBar}), \bibinfo{journal}{Phys. Rev. Lett.}
  \textbf{\bibinfo{volume}{97}}, \bibinfo{pages}{211801}
  (\bibinfo{year}{2006}), \eprint{hep-ex/0607089}.

\bibitem[{\citenamefont{Hokuue et~al.}(2007)}]{Hokuue:2006nr}
\bibinfo{author}{\bibfnamefont{T.}~\bibnamefont{Hokuue}} \bibnamefont{et~al.}
  (\bibinfo{collaboration}{Belle}), \bibinfo{journal}{Phys. Lett.}
  \textbf{\bibinfo{volume}{B648}}, \bibinfo{pages}{139} (\bibinfo{year}{2007}),
  \eprint{hep-ex/0604024}.

\bibitem[{\citenamefont{Adam et~al.}(2007)}]{Adam:2007pv}
\bibinfo{author}{\bibfnamefont{N.~E.} \bibnamefont{Adam}} \bibnamefont{et~al.}
  (\bibinfo{collaboration}{CLEO}), \bibinfo{journal}{Phys. Rev. Lett.}
  \textbf{\bibinfo{volume}{99}}, \bibinfo{pages}{041802}
  (\bibinfo{year}{2007}), \eprint{hep-ex/0703041}.

\bibitem[{\citenamefont{Gray et~al.}(2007)}]{Gray:2007pw}
\bibinfo{author}{\bibfnamefont{R.}~\bibnamefont{Gray}} \bibnamefont{et~al.}
  (\bibinfo{collaboration}{CLEO}), \bibinfo{journal}{Phys. Rev.}
  \textbf{\bibinfo{volume}{D76}}, \bibinfo{pages}{012007}
  (\bibinfo{year}{2007}), \bibinfo{note}{[Addendum: Phys.
  Rev.D76,no.3,039901(2007)]}, \eprint{hep-ex/0703042}.

\bibitem[{\citenamefont{Aubert et~al.}(2008{\natexlab{a}})}]{Aubert:2008bf}
\bibinfo{author}{\bibfnamefont{B.}~\bibnamefont{Aubert}} \bibnamefont{et~al.}
  (\bibinfo{collaboration}{BaBar}), \bibinfo{journal}{Phys. Rev. Lett.}
  \textbf{\bibinfo{volume}{101}}, \bibinfo{pages}{081801}
  (\bibinfo{year}{2008}{\natexlab{a}}), \eprint{0805.2408}.

\bibitem[{\citenamefont{del Amo~Sanchez
  et~al.}(2011{\natexlab{a}})}]{delAmoSanchez:2010af}
\bibinfo{author}{\bibfnamefont{P.}~\bibnamefont{del Amo~Sanchez}}
  \bibnamefont{et~al.} (\bibinfo{collaboration}{BaBar}),
  \bibinfo{journal}{Phys. Rev.} \textbf{\bibinfo{volume}{D83}},
  \bibinfo{pages}{032007} (\bibinfo{year}{2011}{\natexlab{a}}),
  \eprint{1005.3288}.

\bibitem[{\citenamefont{del Amo~Sanchez
  et~al.}(2011{\natexlab{b}})}]{delAmoSanchez:2010zd}
\bibinfo{author}{\bibfnamefont{P.}~\bibnamefont{del Amo~Sanchez}}
  \bibnamefont{et~al.} (\bibinfo{collaboration}{BaBar}),
  \bibinfo{journal}{Phys. Rev.} \textbf{\bibinfo{volume}{D83}},
  \bibinfo{pages}{052011} (\bibinfo{year}{2011}{\natexlab{b}}),
  \eprint{1010.0987}.

\bibitem[{\citenamefont{Ha et~al.}(2011)}]{Ha:2010rf}
\bibinfo{author}{\bibfnamefont{H.}~\bibnamefont{Ha}} \bibnamefont{et~al.}
  (\bibinfo{collaboration}{Belle}), \bibinfo{journal}{Phys. Rev.}
  \textbf{\bibinfo{volume}{D83}}, \bibinfo{pages}{071101}
  (\bibinfo{year}{2011}), \eprint{1012.0090}.

\bibitem[{\citenamefont{Sibidanov et~al.}(2013)}]{Sibidanov:2013rkk}
\bibinfo{author}{\bibfnamefont{A.}~\bibnamefont{Sibidanov}}
  \bibnamefont{et~al.} (\bibinfo{collaboration}{Belle}),
  \bibinfo{journal}{Phys. Rev.} \textbf{\bibinfo{volume}{D88}},
  \bibinfo{pages}{032005} (\bibinfo{year}{2013}), \eprint{1306.2781}.

\bibitem[{\citenamefont{Buskulic et~al.}(1997)}]{Buskulic:1996yq}
\bibinfo{author}{\bibfnamefont{D.}~\bibnamefont{Buskulic}} \bibnamefont{et~al.}
  (\bibinfo{collaboration}{ALEPH}), \bibinfo{journal}{Phys. Lett.}
  \textbf{\bibinfo{volume}{B395}}, \bibinfo{pages}{373} (\bibinfo{year}{1997}).

\bibitem[{\citenamefont{Abbiendi et~al.}(2000)}]{Abbiendi:2000hk}
\bibinfo{author}{\bibfnamefont{G.}~\bibnamefont{Abbiendi}} \bibnamefont{et~al.}
  (\bibinfo{collaboration}{OPAL}), \bibinfo{journal}{Phys. Lett.}
  \textbf{\bibinfo{volume}{B482}}, \bibinfo{pages}{15} (\bibinfo{year}{2000}),
  \eprint{hep-ex/0003013}.

\bibitem[{\citenamefont{Abreu et~al.}(2001)}]{Abreu:2001ic}
\bibinfo{author}{\bibfnamefont{P.}~\bibnamefont{Abreu}} \bibnamefont{et~al.}
  (\bibinfo{collaboration}{DELPHI}), \bibinfo{journal}{Phys. Lett.}
  \textbf{\bibinfo{volume}{B510}}, \bibinfo{pages}{55} (\bibinfo{year}{2001}),
  \eprint{hep-ex/0104026}.

\bibitem[{\citenamefont{Abdallah et~al.}(2004)}]{Abdallah:2004rz}
\bibinfo{author}{\bibfnamefont{J.}~\bibnamefont{Abdallah}} \bibnamefont{et~al.}
  (\bibinfo{collaboration}{DELPHI}), \bibinfo{journal}{Eur. Phys. J.}
  \textbf{\bibinfo{volume}{C33}}, \bibinfo{pages}{213} (\bibinfo{year}{2004}),
  \eprint{hep-ex/0401023}.

\bibitem[{\citenamefont{Adam et~al.}(2003)}]{Adam:2002uw}
\bibinfo{author}{\bibfnamefont{N.~E.} \bibnamefont{Adam}} \bibnamefont{et~al.}
  (\bibinfo{collaboration}{CLEO}), \bibinfo{journal}{Phys. Rev.}
  \textbf{\bibinfo{volume}{D67}}, \bibinfo{pages}{032001}
  (\bibinfo{year}{2003}), \eprint{hep-ex/0210040}.

\bibitem[{\citenamefont{Aubert et~al.}(2008{\natexlab{b}})}]{Aubert:2007rs}
\bibinfo{author}{\bibfnamefont{B.}~\bibnamefont{Aubert}} \bibnamefont{et~al.}
  (\bibinfo{collaboration}{BaBar}), \bibinfo{journal}{Phys. Rev.}
  \textbf{\bibinfo{volume}{D77}}, \bibinfo{pages}{032002}
  (\bibinfo{year}{2008}{\natexlab{b}}), \eprint{0705.4008}.

\bibitem[{\citenamefont{Aubert et~al.}(2008{\natexlab{c}})}]{Aubert:2007qs}
\bibinfo{author}{\bibfnamefont{B.}~\bibnamefont{Aubert}} \bibnamefont{et~al.}
  (\bibinfo{collaboration}{BaBar}), \bibinfo{journal}{Phys. Rev. Lett.}
  \textbf{\bibinfo{volume}{100}}, \bibinfo{pages}{231803}
  (\bibinfo{year}{2008}{\natexlab{c}}), \eprint{0712.3493}.

\bibitem[{\citenamefont{Aubert et~al.}(2009{\natexlab{a}})}]{Aubert:2008yv}
\bibinfo{author}{\bibfnamefont{B.}~\bibnamefont{Aubert}} \bibnamefont{et~al.}
  (\bibinfo{collaboration}{BaBar}), \bibinfo{journal}{Phys. Rev.}
  \textbf{\bibinfo{volume}{D79}}, \bibinfo{pages}{012002}
  (\bibinfo{year}{2009}{\natexlab{a}}), \eprint{0809.0828}.

\bibitem[{\citenamefont{Dungel et~al.}(2010)}]{Dungel:2010uk}
\bibinfo{author}{\bibfnamefont{W.}~\bibnamefont{Dungel}} \bibnamefont{et~al.}
  (\bibinfo{collaboration}{Belle}), \bibinfo{journal}{Phys. Rev.}
  \textbf{\bibinfo{volume}{D82}}, \bibinfo{pages}{112007}
  (\bibinfo{year}{2010}), \eprint{1010.5620}.

\bibitem[{\citenamefont{Aubert et~al.}(2003)}]{Aubert:2003zd}
\bibinfo{author}{\bibfnamefont{B.}~\bibnamefont{Aubert}} \bibnamefont{et~al.}
  (\bibinfo{collaboration}{BaBar}), \bibinfo{journal}{Phys. Rev. Lett.}
  \textbf{\bibinfo{volume}{90}}, \bibinfo{pages}{181801}
  (\bibinfo{year}{2003}), \bibinfo{note}{[eConfC0304052,WG117(2003)]},
  \eprint{hep-ex/0301001}.

\bibitem[{\citenamefont{Schwanda et~al.}(2004)}]{Schwanda:2004fa}
\bibinfo{author}{\bibfnamefont{C.}~\bibnamefont{Schwanda}} \bibnamefont{et~al.}
  (\bibinfo{collaboration}{Belle}), \bibinfo{journal}{Phys. Rev. Lett.}
  \textbf{\bibinfo{volume}{93}}, \bibinfo{pages}{131803}
  (\bibinfo{year}{2004}), \eprint{hep-ex/0402023}.

\bibitem[{\citenamefont{Aubert et~al.}(2009{\natexlab{b}})}]{Aubert:2008ct}
\bibinfo{author}{\bibfnamefont{B.}~\bibnamefont{Aubert}} \bibnamefont{et~al.}
  (\bibinfo{collaboration}{BaBar}), \bibinfo{journal}{Phys. Rev.}
  \textbf{\bibinfo{volume}{D79}}, \bibinfo{pages}{052011}
  (\bibinfo{year}{2009}{\natexlab{b}}), \eprint{0808.3524}.

\bibitem[{\citenamefont{Lees et~al.}(2013{\natexlab{a}})}]{Lees:2012mq}
\bibinfo{author}{\bibfnamefont{J.~P.} \bibnamefont{Lees}} \bibnamefont{et~al.}
  (\bibinfo{collaboration}{BaBar}), \bibinfo{journal}{Phys. Rev.}
  \textbf{\bibinfo{volume}{D87}}, \bibinfo{pages}{032004}
  (\bibinfo{year}{2013}{\natexlab{a}}), \bibinfo{note}{[Erratum: Phys.
  Rev.D87,no.9,099904(2013)]}, \eprint{1205.6245}.

\bibitem[{\citenamefont{Lees et~al.}(2013{\natexlab{b}})}]{Lees:2013gja}
\bibinfo{author}{\bibfnamefont{J.~P.} \bibnamefont{Lees}} \bibnamefont{et~al.}
  (\bibinfo{collaboration}{BaBar}), \bibinfo{journal}{Phys. Rev.}
  \textbf{\bibinfo{volume}{D88}}, \bibinfo{pages}{072006}
  (\bibinfo{year}{2013}{\natexlab{b}}), \eprint{1308.2589}.

\bibitem[{\citenamefont{Dalgic et~al.}(2006)\citenamefont{Dalgic, Gray,
  Wingate, Davies, Lepage, and Shigemitsu}}]{Dalgic:2006dt}
\bibinfo{author}{\bibfnamefont{E.}~\bibnamefont{Dalgic}},
  \bibinfo{author}{\bibfnamefont{A.}~\bibnamefont{Gray}},
  \bibinfo{author}{\bibfnamefont{M.}~\bibnamefont{Wingate}},
  \bibinfo{author}{\bibfnamefont{C.~T.~H.} \bibnamefont{Davies}},
  \bibinfo{author}{\bibfnamefont{G.~P.} \bibnamefont{Lepage}},
  \bibnamefont{and}
  \bibinfo{author}{\bibfnamefont{J.}~\bibnamefont{Shigemitsu}},
  \bibinfo{journal}{Phys. Rev.} \textbf{\bibinfo{volume}{D73}},
  \bibinfo{pages}{074502} (\bibinfo{year}{2006}), \bibinfo{note}{[Erratum:
  Phys. Rev.D75,119906(2007)]}, \eprint{hep-lat/0601021}.

\bibitem[{\citenamefont{Colquhoun et~al.}(2016)\citenamefont{Colquhoun,
  Dowdall, Koponen, Davies, and Lepage}}]{Colquhoun:2015mfa}
\bibinfo{author}{\bibfnamefont{B.}~\bibnamefont{Colquhoun}},
  \bibinfo{author}{\bibfnamefont{R.~J.} \bibnamefont{Dowdall}},
  \bibinfo{author}{\bibfnamefont{J.}~\bibnamefont{Koponen}},
  \bibinfo{author}{\bibfnamefont{C.~T.~H.} \bibnamefont{Davies}},
  \bibnamefont{and} \bibinfo{author}{\bibfnamefont{G.~P.}
  \bibnamefont{Lepage}}, \bibinfo{journal}{Phys. Rev.}
  \textbf{\bibinfo{volume}{D93}}, \bibinfo{pages}{034502}
  (\bibinfo{year}{2016}), \eprint{1510.07446}.

\bibitem[{\citenamefont{Bailey et~al.}(2015{\natexlab{a}})}]{Bailey:2015tia}
\bibinfo{author}{\bibfnamefont{J.~A.} \bibnamefont{Bailey}}
  \bibnamefont{et~al.} (\bibinfo{collaboration}{Fermilab Lattice, MILC}),
  \bibinfo{journal}{Phys. Rev.} \textbf{\bibinfo{volume}{D92}},
  \bibinfo{pages}{014024} (\bibinfo{year}{2015}{\natexlab{a}}),
  \eprint{1503.07839}.

\bibitem[{\citenamefont{Flynn et~al.}(2015)\citenamefont{Flynn, Izubuchi,
  Kawanai, Lehner, Soni, Van~de Water, and Witzel}}]{Flynn:2015mha}
\bibinfo{author}{\bibfnamefont{J.~M.} \bibnamefont{Flynn}},
  \bibinfo{author}{\bibfnamefont{T.}~\bibnamefont{Izubuchi}},
  \bibinfo{author}{\bibfnamefont{T.}~\bibnamefont{Kawanai}},
  \bibinfo{author}{\bibfnamefont{C.}~\bibnamefont{Lehner}},
  \bibinfo{author}{\bibfnamefont{A.}~\bibnamefont{Soni}},
  \bibinfo{author}{\bibfnamefont{R.~S.} \bibnamefont{Van~de Water}},
  \bibnamefont{and} \bibinfo{author}{\bibfnamefont{O.}~\bibnamefont{Witzel}},
  \bibinfo{journal}{Phys. Rev.} \textbf{\bibinfo{volume}{D91}},
  \bibinfo{pages}{074510} (\bibinfo{year}{2015}), \eprint{1501.05373}.

\bibitem[{\citenamefont{Bailey et~al.}(2014)}]{Bailey:2014tva}
\bibinfo{author}{\bibfnamefont{J.~A.} \bibnamefont{Bailey}}
  \bibnamefont{et~al.} (\bibinfo{collaboration}{Fermilab Lattice, MILC}),
  \bibinfo{journal}{Phys. Rev.} \textbf{\bibinfo{volume}{D89}},
  \bibinfo{pages}{114504} (\bibinfo{year}{2014}), \eprint{1403.0635}.

\bibitem[{\citenamefont{Na et~al.}(2015)\citenamefont{Na, Bouchard, Lepage,
  Monahan, and Shigemitsu}}]{Na:2015kha}
\bibinfo{author}{\bibfnamefont{H.}~\bibnamefont{Na}},
  \bibinfo{author}{\bibfnamefont{C.~M.} \bibnamefont{Bouchard}},
  \bibinfo{author}{\bibfnamefont{G.~P.} \bibnamefont{Lepage}},
  \bibinfo{author}{\bibfnamefont{C.}~\bibnamefont{Monahan}}, \bibnamefont{and}
  \bibinfo{author}{\bibfnamefont{J.}~\bibnamefont{Shigemitsu}}
  (\bibinfo{collaboration}{HPQCD}), \bibinfo{journal}{Phys. Rev.}
  \textbf{\bibinfo{volume}{D92}}, \bibinfo{pages}{054510}
  (\bibinfo{year}{2015}), \bibinfo{note}{[Erratum: Phys.
  Rev.D93,no.11,119906(2016)]}, \eprint{1505.03925}.

\bibitem[{\citenamefont{Bailey et~al.}(2015{\natexlab{b}})}]{Lattice:2015rga}
\bibinfo{author}{\bibfnamefont{J.~A.} \bibnamefont{Bailey}}
  \bibnamefont{et~al.} (\bibinfo{collaboration}{MILC}), \bibinfo{journal}{Phys.
  Rev.} \textbf{\bibinfo{volume}{D92}}, \bibinfo{pages}{034506}
  (\bibinfo{year}{2015}{\natexlab{b}}), \eprint{1503.07237}.

\bibitem[{\citenamefont{Harrison et~al.}(2018)\citenamefont{Harrison, Davies,
  and Wingate}}]{Harrison:2017fmw}
\bibinfo{author}{\bibfnamefont{J.}~\bibnamefont{Harrison}},
  \bibinfo{author}{\bibfnamefont{C.}~\bibnamefont{Davies}}, \bibnamefont{and}
  \bibinfo{author}{\bibfnamefont{M.}~\bibnamefont{Wingate}}
  (\bibinfo{collaboration}{HPQCD}), \bibinfo{journal}{Phys. Rev.}
  \textbf{\bibinfo{volume}{D97}}, \bibinfo{pages}{054502}
  (\bibinfo{year}{2018}), \eprint{1711.11013}.

\bibitem[{\citenamefont{Ball and Zwicky}(2005)}]{Ball:2004ye}
\bibinfo{author}{\bibfnamefont{P.}~\bibnamefont{Ball}} \bibnamefont{and}
  \bibinfo{author}{\bibfnamefont{R.}~\bibnamefont{Zwicky}},
  \bibinfo{journal}{Phys. Rev.} \textbf{\bibinfo{volume}{D71}},
  \bibinfo{pages}{014015} (\bibinfo{year}{2005}), \eprint{hep-ph/0406232}.

\bibitem[{\citenamefont{Duplancic et~al.}(2008)\citenamefont{Duplancic,
  Khodjamirian, Mannel, Melic, and Offen}}]{Duplancic:2008ix}
\bibinfo{author}{\bibfnamefont{G.}~\bibnamefont{Duplancic}},
  \bibinfo{author}{\bibfnamefont{A.}~\bibnamefont{Khodjamirian}},
  \bibinfo{author}{\bibfnamefont{T.}~\bibnamefont{Mannel}},
  \bibinfo{author}{\bibfnamefont{B.}~\bibnamefont{Melic}}, \bibnamefont{and}
  \bibinfo{author}{\bibfnamefont{N.}~\bibnamefont{Offen}},
  \bibinfo{journal}{JHEP} \textbf{\bibinfo{volume}{04}}, \bibinfo{pages}{014}
  (\bibinfo{year}{2008}), \eprint{0801.1796}.

\bibitem[{\citenamefont{Khodjamirian et~al.}(2011)\citenamefont{Khodjamirian,
  Mannel, Offen, and Wang}}]{Khodjamirian:2011ub}
\bibinfo{author}{\bibfnamefont{A.}~\bibnamefont{Khodjamirian}},
  \bibinfo{author}{\bibfnamefont{T.}~\bibnamefont{Mannel}},
  \bibinfo{author}{\bibfnamefont{N.}~\bibnamefont{Offen}}, \bibnamefont{and}
  \bibinfo{author}{\bibfnamefont{Y.~M.} \bibnamefont{Wang}},
  \bibinfo{journal}{Phys. Rev.} \textbf{\bibinfo{volume}{D83}},
  \bibinfo{pages}{094031} (\bibinfo{year}{2011}), \eprint{1103.2655}.

\bibitem[{\citenamefont{Bharucha}(2012)}]{Bharucha:2012wy}
\bibinfo{author}{\bibfnamefont{A.}~\bibnamefont{Bharucha}},
  \bibinfo{journal}{JHEP} \textbf{\bibinfo{volume}{05}}, \bibinfo{pages}{092}
  (\bibinfo{year}{2012}), \eprint{1203.1359}.

\bibitem[{\citenamefont{Sentitemsu~Imsong
  et~al.}(2015)\citenamefont{Sentitemsu~Imsong, Khodjamirian, Mannel, and van
  Dyk}}]{Imsong:2014oqa}
\bibinfo{author}{\bibfnamefont{I.}~\bibnamefont{Sentitemsu~Imsong}},
  \bibinfo{author}{\bibfnamefont{A.}~\bibnamefont{Khodjamirian}},
  \bibinfo{author}{\bibfnamefont{T.}~\bibnamefont{Mannel}}, \bibnamefont{and}
  \bibinfo{author}{\bibfnamefont{D.}~\bibnamefont{van Dyk}},
  \bibinfo{journal}{JHEP} \textbf{\bibinfo{volume}{02}}, \bibinfo{pages}{126}
  (\bibinfo{year}{2015}), \eprint{1409.7816}.

\bibitem[{\citenamefont{Bigi et~al.}(1995)\citenamefont{Bigi, Shifman,
  Uraltsev, and Vainshtein}}]{Bigi:1994ga}
\bibinfo{author}{\bibfnamefont{I.~I.~Y.} \bibnamefont{Bigi}},
  \bibinfo{author}{\bibfnamefont{M.~A.} \bibnamefont{Shifman}},
  \bibinfo{author}{\bibfnamefont{N.~G.} \bibnamefont{Uraltsev}},
  \bibnamefont{and} \bibinfo{author}{\bibfnamefont{A.~I.}
  \bibnamefont{Vainshtein}}, \bibinfo{journal}{Phys. Rev.}
  \textbf{\bibinfo{volume}{D52}}, \bibinfo{pages}{196} (\bibinfo{year}{1995}),
  \eprint{hep-ph/9405410}.

\bibitem[{\citenamefont{Kapustin et~al.}(1996)\citenamefont{Kapustin, Ligeti,
  Wise, and Grinstein}}]{Kapustin:1996dy}
\bibinfo{author}{\bibfnamefont{A.}~\bibnamefont{Kapustin}},
  \bibinfo{author}{\bibfnamefont{Z.}~\bibnamefont{Ligeti}},
  \bibinfo{author}{\bibfnamefont{M.~B.} \bibnamefont{Wise}}, \bibnamefont{and}
  \bibinfo{author}{\bibfnamefont{B.}~\bibnamefont{Grinstein}},
  \bibinfo{journal}{Phys. Lett.} \textbf{\bibinfo{volume}{B375}},
  \bibinfo{pages}{327} (\bibinfo{year}{1996}), \eprint{hep-ph/9602262}.

\bibitem[{\citenamefont{Gambino et~al.}(2010)\citenamefont{Gambino, Mannel, and
  Uraltsev}}]{Gambino:2010bp}
\bibinfo{author}{\bibfnamefont{P.}~\bibnamefont{Gambino}},
  \bibinfo{author}{\bibfnamefont{T.}~\bibnamefont{Mannel}}, \bibnamefont{and}
  \bibinfo{author}{\bibfnamefont{N.}~\bibnamefont{Uraltsev}},
  \bibinfo{journal}{Phys. Rev.} \textbf{\bibinfo{volume}{D81}},
  \bibinfo{pages}{113002} (\bibinfo{year}{2010}), \eprint{1004.2859}.

\bibitem[{\citenamefont{Gambino et~al.}(2012)\citenamefont{Gambino, Mannel, and
  Uraltsev}}]{Gambino:2012rd}
\bibinfo{author}{\bibfnamefont{P.}~\bibnamefont{Gambino}},
  \bibinfo{author}{\bibfnamefont{T.}~\bibnamefont{Mannel}}, \bibnamefont{and}
  \bibinfo{author}{\bibfnamefont{N.}~\bibnamefont{Uraltsev}},
  \bibinfo{journal}{JHEP} \textbf{\bibinfo{volume}{10}}, \bibinfo{pages}{169}
  (\bibinfo{year}{2012}), \eprint{1206.2296}.

\bibitem[{\citenamefont{Aaij et~al.}(2015)}]{Aaij:2015bfa}
\bibinfo{author}{\bibfnamefont{R.}~\bibnamefont{Aaij}} \bibnamefont{et~al.}
  (\bibinfo{collaboration}{LHCb}), \bibinfo{journal}{Nature Phys.}
  \textbf{\bibinfo{volume}{11}}, \bibinfo{pages}{743} (\bibinfo{year}{2015}),
  \eprint{1504.01568}.

\bibitem[{\citenamefont{Aaij et~al.}(2017)}]{Aaij:2017svr}
\bibinfo{author}{\bibfnamefont{R.}~\bibnamefont{Aaij}} \bibnamefont{et~al.}
  (\bibinfo{collaboration}{LHCb}), \bibinfo{journal}{Phys. Rev.}
  \textbf{\bibinfo{volume}{D96}}, \bibinfo{pages}{112005}
  (\bibinfo{year}{2017}), \eprint{1709.01920}.

\bibitem[{\citenamefont{Detmold et~al.}(2015)\citenamefont{Detmold, Lehner, and
  Meinel}}]{Detmold:2015aaa}
\bibinfo{author}{\bibfnamefont{W.}~\bibnamefont{Detmold}},
  \bibinfo{author}{\bibfnamefont{C.}~\bibnamefont{Lehner}}, \bibnamefont{and}
  \bibinfo{author}{\bibfnamefont{S.}~\bibnamefont{Meinel}},
  \bibinfo{journal}{Phys. Rev.} \textbf{\bibinfo{volume}{D92}},
  \bibinfo{pages}{034503} (\bibinfo{year}{2015}), \eprint{1503.01421}.

\bibitem[{\citenamefont{Bouchard et~al.}(2014)\citenamefont{Bouchard, Lepage,
  Monahan, Na, and Shigemitsu}}]{Bouchard:2014ypa}
\bibinfo{author}{\bibfnamefont{C.~M.} \bibnamefont{Bouchard}},
  \bibinfo{author}{\bibfnamefont{G.~P.} \bibnamefont{Lepage}},
  \bibinfo{author}{\bibfnamefont{C.}~\bibnamefont{Monahan}},
  \bibinfo{author}{\bibfnamefont{H.}~\bibnamefont{Na}}, \bibnamefont{and}
  \bibinfo{author}{\bibfnamefont{J.}~\bibnamefont{Shigemitsu}},
  \bibinfo{journal}{Phys. Rev.} \textbf{\bibinfo{volume}{D90}},
  \bibinfo{pages}{054506} (\bibinfo{year}{2014}).

\bibitem[{\citenamefont{Monahan et~al.}(2017)\citenamefont{Monahan, Na,
  Bouchard, Lepage, and Shigemitsu}}]{Monahan:2017uby}
\bibinfo{author}{\bibfnamefont{C.~J.} \bibnamefont{Monahan}},
  \bibinfo{author}{\bibfnamefont{H.}~\bibnamefont{Na}},
  \bibinfo{author}{\bibfnamefont{C.~M.} \bibnamefont{Bouchard}},
  \bibinfo{author}{\bibfnamefont{G.~P.} \bibnamefont{Lepage}},
  \bibnamefont{and}
  \bibinfo{author}{\bibfnamefont{J.}~\bibnamefont{Shigemitsu}},
  \bibinfo{journal}{Phys. Rev.} \textbf{\bibinfo{volume}{D95}},
  \bibinfo{pages}{114506} (\bibinfo{year}{2017}), \eprint{1703.09728}.

\bibitem[{\citenamefont{Bazavov et~al.}(2010)}]{Bazavov:2009bb}
\bibinfo{author}{\bibfnamefont{A.}~\bibnamefont{Bazavov}} \bibnamefont{et~al.}
  (\bibinfo{collaboration}{MILC}), \bibinfo{journal}{Rev. Mod. Phys.}
  \textbf{\bibinfo{volume}{82}}, \bibinfo{pages}{1349} (\bibinfo{year}{2010}).

\bibitem[{\citenamefont{Bernard}(2002)}]{Bernard:2001yj}
\bibinfo{author}{\bibfnamefont{C.}~\bibnamefont{Bernard}}
  (\bibinfo{collaboration}{MILC}), \bibinfo{journal}{Phys. Rev.}
  \textbf{\bibinfo{volume}{D65}}, \bibinfo{pages}{054031}
  (\bibinfo{year}{2002}), \eprint{hep-lat/0111051}.

\bibitem[{\citenamefont{Na et~al.}(2012)\citenamefont{Na, Monahan, Davies,
  Horgan, Lepage, and Shigemitsu}}]{Na:2012kp}
\bibinfo{author}{\bibfnamefont{H.}~\bibnamefont{Na}},
  \bibinfo{author}{\bibfnamefont{C.~J.} \bibnamefont{Monahan}},
  \bibinfo{author}{\bibfnamefont{C.~T.} \bibnamefont{Davies}},
  \bibinfo{author}{\bibfnamefont{R.}~\bibnamefont{Horgan}},
  \bibinfo{author}{\bibfnamefont{G.~P.} \bibnamefont{Lepage}},
  \bibnamefont{and}
  \bibinfo{author}{\bibfnamefont{J.}~\bibnamefont{Shigemitsu}},
  \bibinfo{journal}{Phys.Rev.} \textbf{\bibinfo{volume}{D86}},
  \bibinfo{pages}{034506} (\bibinfo{year}{2012}).

\bibitem[{\citenamefont{Monahan et~al.}(2013)\citenamefont{Monahan, Shigemitsu,
  and Horgan}}]{Monahan:2012dq}
\bibinfo{author}{\bibfnamefont{C.}~\bibnamefont{Monahan}},
  \bibinfo{author}{\bibfnamefont{J.}~\bibnamefont{Shigemitsu}},
  \bibnamefont{and} \bibinfo{author}{\bibfnamefont{R.}~\bibnamefont{Horgan}},
  \bibinfo{journal}{Phys.Rev.} \textbf{\bibinfo{volume}{D87}},
  \bibinfo{pages}{034017} (\bibinfo{year}{2013}).

\bibitem[{\citenamefont{Na et~al.}(2010)\citenamefont{Na, Davies, Follana,
  Lepage, and Shigemitsu}}]{Na:2010uf}
\bibinfo{author}{\bibfnamefont{H.}~\bibnamefont{Na}},
  \bibinfo{author}{\bibfnamefont{C.~T.} \bibnamefont{Davies}},
  \bibinfo{author}{\bibfnamefont{E.}~\bibnamefont{Follana}},
  \bibinfo{author}{\bibfnamefont{G.~P.} \bibnamefont{Lepage}},
  \bibnamefont{and}
  \bibinfo{author}{\bibfnamefont{J.}~\bibnamefont{Shigemitsu}},
  \bibinfo{journal}{Phys.Rev.} \textbf{\bibinfo{volume}{D82}},
  \bibinfo{pages}{114506} (\bibinfo{year}{2010}).

\bibitem[{\citenamefont{{Lepage, G.P.}}({\natexlab{a}})}]{lsqfit}
\bibinfo{author}{\bibnamefont{{Lepage, G.P.}}}, \emph{\bibinfo{title}{lsqfit
  v4.8.5.1}}, \urlprefix\url{https://doi.org/10.5281/zenodo.10236}.

\bibitem[{\citenamefont{{Lepage, G.P.}}({\natexlab{b}})}]{corrfitter}
\bibinfo{author}{\bibnamefont{{Lepage, G.P.}}},
  \emph{\bibinfo{title}{corrfitter v3.7.1}},
  \urlprefix\url{https://doi.org/10.5281/zenodo.10237}.

\bibitem[{\citenamefont{Na et~al.}(2011)\citenamefont{Na, Davies, Follana,
  Koponen, Lepage et~al.}}]{Na:2011mc}
\bibinfo{author}{\bibfnamefont{H.}~\bibnamefont{Na}},
  \bibinfo{author}{\bibfnamefont{C.~T.} \bibnamefont{Davies}},
  \bibinfo{author}{\bibfnamefont{E.}~\bibnamefont{Follana}},
  \bibinfo{author}{\bibfnamefont{J.}~\bibnamefont{Koponen}},
  \bibinfo{author}{\bibfnamefont{G.~P.} \bibnamefont{Lepage}},
  \bibnamefont{et~al.}, \bibinfo{journal}{Phys.Rev.}
  \textbf{\bibinfo{volume}{D84}}, \bibinfo{pages}{114505}
  (\bibinfo{year}{2011}).

\bibitem[{\citenamefont{Bouchard
  et~al.}(2013{\natexlab{a}})\citenamefont{Bouchard, Lepage, Monahan, Na, and
  Shigemitsu}}]{Bouchard:2013mia}
\bibinfo{author}{\bibfnamefont{C.}~\bibnamefont{Bouchard}},
  \bibinfo{author}{\bibfnamefont{G.~P.} \bibnamefont{Lepage}},
  \bibinfo{author}{\bibfnamefont{C.}~\bibnamefont{Monahan}},
  \bibinfo{author}{\bibfnamefont{H.}~\bibnamefont{Na}}, \bibnamefont{and}
  \bibinfo{author}{\bibfnamefont{J.}~\bibnamefont{Shigemitsu}}
  (\bibinfo{collaboration}{HPQCD}), \bibinfo{journal}{Phys. Rev. Lett.}
  \textbf{\bibinfo{volume}{111}}, \bibinfo{pages}{162002}
  (\bibinfo{year}{2013}{\natexlab{a}}), \bibinfo{note}{[Erratum: Phys. Rev.
  Lett.112,no.14,149902(2014)]}.

\bibitem[{\citenamefont{Bouchard
  et~al.}(2013{\natexlab{b}})\citenamefont{Bouchard, Lepage, Monahan, Na, and
  Shigemitsu}}]{Bouchard:2013pna}
\bibinfo{author}{\bibfnamefont{C.}~\bibnamefont{Bouchard}},
  \bibinfo{author}{\bibfnamefont{G.~P.} \bibnamefont{Lepage}},
  \bibinfo{author}{\bibfnamefont{C.}~\bibnamefont{Monahan}},
  \bibinfo{author}{\bibfnamefont{H.}~\bibnamefont{Na}}, \bibnamefont{and}
  \bibinfo{author}{\bibfnamefont{J.}~\bibnamefont{Shigemitsu}}
  (\bibinfo{collaboration}{HPQCD}), \bibinfo{journal}{Phys. Rev.}
  \textbf{\bibinfo{volume}{D88}}, \bibinfo{pages}{054509}
  (\bibinfo{year}{2013}{\natexlab{b}}), \bibinfo{note}{[Erratum: Phys.
  Rev.D88,no.7,079901(2013)]}.

\bibitem[{\citenamefont{Bourrely et~al.}(2009)\citenamefont{Bourrely, Caprini,
  and Lellouch}}]{Bourrely:2008za}
\bibinfo{author}{\bibfnamefont{C.}~\bibnamefont{Bourrely}},
  \bibinfo{author}{\bibfnamefont{I.}~\bibnamefont{Caprini}}, \bibnamefont{and}
  \bibinfo{author}{\bibfnamefont{L.}~\bibnamefont{Lellouch}},
  \bibinfo{journal}{Phys. Rev.} \textbf{\bibinfo{volume}{D79}},
  \bibinfo{pages}{013008} (\bibinfo{year}{2009}), \bibinfo{note}{[Erratum:
  Phys. Rev.D82,099902(2010)]}.

\bibitem[{\citenamefont{Gregory et~al.}(2010)\citenamefont{Gregory, Davies,
  Follana, Gamiz, Kendall, Lepage, Na, Shigemitsu, and Wong}}]{Gregory:2009hq}
\bibinfo{author}{\bibfnamefont{E.~B.} \bibnamefont{Gregory}},
  \bibinfo{author}{\bibfnamefont{C.~T.~H.} \bibnamefont{Davies}},
  \bibinfo{author}{\bibfnamefont{E.}~\bibnamefont{Follana}},
  \bibinfo{author}{\bibfnamefont{E.}~\bibnamefont{Gamiz}},
  \bibinfo{author}{\bibfnamefont{I.~D.} \bibnamefont{Kendall}},
  \bibinfo{author}{\bibfnamefont{G.~P.} \bibnamefont{Lepage}},
  \bibinfo{author}{\bibfnamefont{H.}~\bibnamefont{Na}},
  \bibinfo{author}{\bibfnamefont{J.}~\bibnamefont{Shigemitsu}},
  \bibnamefont{and} \bibinfo{author}{\bibfnamefont{K.~Y.} \bibnamefont{Wong}},
  \bibinfo{journal}{Phys. Rev. Lett.} \textbf{\bibinfo{volume}{104}},
  \bibinfo{pages}{022001} (\bibinfo{year}{2010}).

\bibitem[{\citenamefont{Bijnens and Jemos}(2010)}]{Bijnens:2010ws}
\bibinfo{author}{\bibfnamefont{J.}~\bibnamefont{Bijnens}} \bibnamefont{and}
  \bibinfo{author}{\bibfnamefont{I.}~\bibnamefont{Jemos}},
  \bibinfo{journal}{Nucl. Phys.} \textbf{\bibinfo{volume}{B840}},
  \bibinfo{pages}{54} (\bibinfo{year}{2010}), \bibinfo{note}{[Erratum: Nucl.
  Phys.B844,182(2011)]}, \eprint{1006.1197}.

\bibitem[{\citenamefont{Bijnens and Jemos}(2011)}]{Bijnens:2010jg}
\bibinfo{author}{\bibfnamefont{J.}~\bibnamefont{Bijnens}} \bibnamefont{and}
  \bibinfo{author}{\bibfnamefont{I.}~\bibnamefont{Jemos}},
  \bibinfo{journal}{Nucl. Phys.} \textbf{\bibinfo{volume}{B846}},
  \bibinfo{pages}{145} (\bibinfo{year}{2011}), \eprint{1011.6531}.

\bibitem[{\citenamefont{Lubicz et~al.}(2016)\citenamefont{Lubicz, Riggio,
  Salerno, Simula, and Tarantino}}]{Lubicz:2016wwx}
\bibinfo{author}{\bibfnamefont{V.}~\bibnamefont{Lubicz}},
  \bibinfo{author}{\bibfnamefont{L.}~\bibnamefont{Riggio}},
  \bibinfo{author}{\bibfnamefont{G.}~\bibnamefont{Salerno}},
  \bibinfo{author}{\bibfnamefont{S.}~\bibnamefont{Simula}}, \bibnamefont{and}
  \bibinfo{author}{\bibfnamefont{C.}~\bibnamefont{Tarantino}},
  \bibinfo{journal}{PoS} \textbf{\bibinfo{volume}{LATTICE2016}},
  \bibinfo{pages}{280} (\bibinfo{year}{2016}), \eprint{1611.00022}.

\bibitem[{\citenamefont{Lees et~al.}(2012)}]{Lees:2012xj}
\bibinfo{author}{\bibfnamefont{J.~P.} \bibnamefont{Lees}} \bibnamefont{et~al.}
  (\bibinfo{collaboration}{BaBar}), \bibinfo{journal}{Phys. Rev. Lett.}
  \textbf{\bibinfo{volume}{109}}, \bibinfo{pages}{101802}
  (\bibinfo{year}{2012}).

\bibitem[{\citenamefont{Lees et~al.}(2013{\natexlab{c}})}]{Lees:2013uzd}
\bibinfo{author}{\bibfnamefont{J.~P.} \bibnamefont{Lees}} \bibnamefont{et~al.}
  (\bibinfo{collaboration}{BaBar}), \bibinfo{journal}{Phys. Rev.}
  \textbf{\bibinfo{volume}{D88}}, \bibinfo{pages}{072012}
  (\bibinfo{year}{2013}{\natexlab{c}}).

\bibitem[{\citenamefont{Huschle et~al.}(2015)}]{Huschle:2015rga}
\bibinfo{author}{\bibfnamefont{M.}~\bibnamefont{Huschle}} \bibnamefont{et~al.}
  (\bibinfo{collaboration}{Belle}), \bibinfo{journal}{Phys. Rev.}
  \textbf{\bibinfo{volume}{D92}}, \bibinfo{pages}{072014}
  (\bibinfo{year}{2015}).

\bibitem[{\citenamefont{Amhis et~al.}(2016)}]{hfag:2016rds}
\bibinfo{author}{\bibfnamefont{Y.}~\bibnamefont{Amhis}} \bibnamefont{et~al.}
  (\bibinfo{collaboration}{HFAG}), \emph{\bibinfo{title}{{Average of $R(D)$ and
  $R(D^\ast)$}}} (\bibinfo{year}{2016}),
  \urlprefix\url{https://www.slac.stanford.edu/xorg/hfag/semi/eps15/eps15_dtaunu.html}.

\bibitem[{\citenamefont{Kamenik and Mescia}(2008)}]{Kamenik:2008tj}
\bibinfo{author}{\bibfnamefont{J.~F.} \bibnamefont{Kamenik}} \bibnamefont{and}
  \bibinfo{author}{\bibfnamefont{F.}~\bibnamefont{Mescia}},
  \bibinfo{journal}{Phys. Rev.} \textbf{\bibinfo{volume}{D78}},
  \bibinfo{pages}{014003} (\bibinfo{year}{2008}).

\bibitem[{\citenamefont{Bailey et~al.}(2012)\citenamefont{Bailey, Bazavov,
  Bernard, Bouchard, DeTar et~al.}}]{Bailey:2012rr}
\bibinfo{author}{\bibfnamefont{J.~A.} \bibnamefont{Bailey}},
  \bibinfo{author}{\bibfnamefont{A.}~\bibnamefont{Bazavov}},
  \bibinfo{author}{\bibfnamefont{C.}~\bibnamefont{Bernard}},
  \bibinfo{author}{\bibfnamefont{C.}~\bibnamefont{Bouchard}},
  \bibinfo{author}{\bibfnamefont{C.}~\bibnamefont{DeTar}},
  \bibnamefont{et~al.}, \bibinfo{journal}{Phys.Rev.}
  \textbf{\bibinfo{volume}{D85}}, \bibinfo{pages}{114502}
  (\bibinfo{year}{2012}).

\end{thebibliography}


\end{document}